\documentclass[fleqn,usenatbib]{mnras}

\IfFileExists{newtxtext.sty}{\usepackage{newtxtext,newtxmath}}{\usepackage{mathptmx}\usepackage{amssymb}}

\usepackage[T1]{fontenc}

\DeclareRobustCommand{\VAN}[3]{#2}
\let\VANthebibliography\thebibliography
\def\thebibliography{\DeclareRobustCommand{\VAN}[3]{##3}\VANthebibliography}

\usepackage{graphicx}	% Including figure files
\usepackage{amsmath}	% Advanced maths commands
\usepackage{natbib}
\usepackage{docmute}
\usepackage{threeparttable}
\usepackage{pifont}
\usepackage{multirow}
\usepackage{caption}
\usepackage{soul}
\usepackage[table]{xcolor}
\usepackage{lastpage}
\usepackage{hyperref}
\usepackage{xspace}
\newif\ifshowalttext
\showalttextfalse
\newcommand{\alttext}[1]{%
  \ifshowalttext
    \par\smallskip\noindent\textbf{Alt text:} #1\par
  \fi
}
\ifshowalttext
  \setkeys{Gin}{height=0.6\textheight,keepaspectratio}
\fi

\defcitealias{collaborationPlanck2018Results2020}{Planck Collaboration 2020}
\newcommand{\nc}{\texttt{K26}\xspace}
\newcommand{\fid}{\texttt{FID}\xspace}
\newcommand{\norad}{\texttt{NoRad}\xspace}
\newcommand{\hires}{\texttt{HiRes}\xspace}

\usepackage[normalem]{ulem}

\newcommand{\Kang}{%
  \hyperlink{cite.kangMitigatingOvercoolingProblem2026}
  {K26}\xspace
}
\newcommand{\Kangfive}{%
  \hyperlink{cite.kangImpactStarFormation2025}
  {K25}\xspace
}
\title[Star clusters at cosmic dawn]{Star clusters at cosmic dawn I: the emergence of the initial cluster mass function}
\author[F. S. Thompson et al.]{
Fred S. Thompson,$^{1, 2}$\thanks{E-mail: frederick.thompson@physics.ox.ac.uk}
Julien Devriendt,$^{1}$
Adrianne Slyz$,^{1}$
Taysun Kimm,$^{2}$
and Cheonsu Kang$^{2}$
\\
$^{1}$Sub-department of Astrophysics, University of Oxford, Denys Wilkinson Building, Keble Road, Oxford OX1 3RH, UK\\
$^{2}$Department of Astronomy, Yonsei University, 50 Yonsei-ro, Seodaemun-gu, Seoul 03722, Republic of Korea
}

\date{Accepted XXX. Received YYY; in original form ZZZ}

\pubyear{\the\year{}}

\begin{document}
\label{firstpage}
\pagerange{\pageref{firstpage}--\pageref{LastPage}}
\maketitle

\begin{abstract}
\textit{JWST} observations of lensed galaxies reveal unusually bright star-cluster candidates within the first billion years after the Big Bang. However, the physical processes shaping the initial cluster mass function (ICMF) remain uncertain. We study the formation of a galaxy embedded in a dark matter halo whose virial mass reaches $10^{10}\,{\rm M_\odot}$ at $z=6$, using cosmological radiation-hydrodynamics zoom-in simulations. Its star clusters, represented by sink particles, accrete gas and form stellar populations that supply radiation and supernova feedback, thereby regulating their own growth.
As we do not prescribe an efficiency per free-fall time or a fixed star-cluster formation duration, the ICMF emerges self-consistently. We assess the effects of numerical resolution and stellar radiation. Schechter fits to the central-galaxy ICMF give truncation masses near $6\times10^6\,{\rm M_\odot}$, with substantial uncertainty.
At fixed resolution, including stellar radiation makes the low-mass slope of the ICMF shallower, from $-1.60$ to $-1.34$, while doubling the spatial resolution steepens it back to $-1.54$.
Stellar radiation also reduces the abundance of low-mass clouds, particularly those with lower gas densities, and shortens the formation time of star clusters to about a Myr, i.e. well before their first supernovae go off.
On average, more rapidly fed star clusters accrete for longer, making their final masses rise faster than linearly with the initial accretion rate and contributing to a flatter ICMF.
At fixed host-galaxy star formation rate, our brightest simulated clusters reach UV luminosities comparable to those of observed lensed star-cluster candidates.
\end{abstract}


%%%%%%%%%%%%%%%%%%%%%%%%%%%%%%%%%%%%%%%%%%%%%%%%%%

%%%%%%%%%%%%%%%%% BODY OF PAPER %%%%%%%%%%%%%%%%%%

\section{Introduction}
\label{sec:intro}

Stars form in hierarchical structures within molecular clouds, with an environmentally dependent fraction remaining in bound star clusters \citep{ladaEmbeddedClustersMolecular2003, portegieszwartYoungMassiveStar2010, bressertSpatialDistributionStar2010, grashaHierarchicalDistributionYoung2017}. The cluster formation efficiency, $\Gamma$, measures the fraction of star formation that ends up in bound clusters and is found to rise with the gas pressure and star formation rate surface density of the host \citep{goddardFractionStarClusters2010, kruijssenFractionStarFormation2012, adamoProbingRoleGalactic2015, adamoStarClustersFar2020}. The distribution of cluster birth masses determines how stellar feedback is concentrated because radiation and supernovae (SNe) from coeval massive stars can overlap in space and time, changing how their energy couples to the interstellar medium (ISM) \citep{murrayDisruptionGiantMolecular2010, gentryEnhancedMomentumFeedback2017, fieldingClusteredSupernovaeDrive2018}. It also supplies the initial conditions for the clusters that survive to the present day \citep{pfefferEMOSAICSProjectSimulating2018, kruijssenFormationAssemblyHistory2019, chenCatalogueModelStar2024}.

Moreover, the youngest and most massive of these clusters are bright enough to be observed individually in the early Universe.
Indeed, \textit{JWST} observations combined with gravitational lensing now allow us to probe their formation. Magnifications of tens to hundreds resolve individual star clusters and compact clumps in galaxies from $z\simeq6$ to beyond $z=10$ \citep[e.g.][]{vanzellaJWSTNIRCamProbes2023, mowlaFormationLowmassGalaxy2024, adamoBoundStarClusters2024a, messaAnatomyZ6Lymana2025, fujimotoPrimordialRotatingDisk2025, bradacStarFormationCosmic2025, abdurroufSpatiallyResolvedPhysical2025, nakaneVENUSStronglyLensed2025}. In the Cosmic Gems arc at $z=9.625$ \citep{messaJWSTSpectroscopicConfirmation2026}, for example, \citet{adamoBoundStarClusters2024a} find five clusters younger than 50\,Myr with masses of $\sim10^6\,{\rm M_\odot}$. These objects appear unusually bright for their host galaxies. In local galaxies, the luminosity of the brightest young cluster rises with the star formation rate (SFR) of its host \citep{larsenLuminosityFunctionStar2002, weidnerImplicationsFormationStar2004, bastianStarFormationRate2008}. At high redshift, \citet{pfefferComparingEMOSAICSPredictions2025} find that the brightest cluster candidates in lensed galaxies are $\approx$1--1.5\,mag brighter and $\approx$0.5\,dex more massive than the median prediction of E-MOSAICS, which draws clusters from a power-law ICMF of slope $-2$ with an environmentally set truncation. The brightest cluster at a given SFR is a useful probe of the upper end of the ICMF \citep{bastianStarFormationRate2008, adamoProbingRoleGalactic2015}, and at these redshifts it is one of the few population statistics that can be measured. Why the first clusters should be so bright is an open question. One possibility, the so-called `feedback-free' scenario, is that early star-forming clouds are dense enough to convert much of their gas into stars before feedback can act \citep{dekelEfficientFormationMassive2023a, liFeedbackFreeStarburstsCosmic2024}.

In the local Universe, clouds and young clusters have mass functions close to ${\rm d}N/{\rm d}M\propto M^{-2}$ \citep{portegieszwartYoungMassiveStar2010, krumholzStarClustersCosmic2019a, adamoStarClustersFar2020}, as expected from fragmentation of a scale-free, turbulent ISM \citep{elmegreenUniversalFormationMechanism1997, guszejnovUniversalScalingRelations2018}. A direct measurement of the cluster mass function at an earlier epoch is consistent with this form, giving a pure power-law slope of $-1.89^{+0.13}_{-0.12}$ \citep{claeyssensFirstGLIMPSEStar2026}, whereas local mergers and luminous infrared galaxies have shallower slopes, down to about $-1.4$ \citep{adamoStarClusterFormation2020, lindenMassiveStarCluster2021}. Similar departures occur in simulations of starbursts in merging dwarfs \citep{lahenFormationLowMetallicity2019} and in dwarf galaxies before reionization, shortly after a starburst \citep{garciaStarClusterFormation2023, garciaSeedingCoresPathway2025a}. These shallower slopes can arise if the star formation efficiency, or the fraction of stars that remains bound, rises with cloud density or mass. The most massive clouds then convert more of their gas into bound clusters, shifting the cluster population towards higher masses \citep{fallStellarFeedbackMolecular2010, liStarClusterFormation2018, grudicGreatBallsFIRE2023}.

However, the ICMF often departs from a pure power law at its high-mass end. In nearby spiral galaxies it is truncated around $\sim10^5\,{\rm M_\odot}$ \citep{gielesLuminosityFunctionYoung2006, adamoProbingRoleGalactic2015}, whereas in merging and luminous infrared galaxies the truncation can reach $\sim10^7\,{\rm M_\odot}$ \citep{randriamanakotoYoungMassiveClusters2019, adamoStarClusterFormation2020, lindenMassiveStarCluster2021}. This environmental dependence suggests that the upper mass scale reflects the gas available for collapse. Analytic models set it by the largest mass that can collapse against galactic shear, or by the fraction of that mass that collapses before stellar feedback disperses the gas \citep{reina-camposUnifiedModelMaximum2017a}, consistent with the observed rise of cluster truncation mass with the host galaxy star formation rate surface density \citep{johnsonPanchromaticHubbleAndromeda2017}. At high redshift, the lensed sample of \citet{claeyssensFirstGLIMPSEStar2026} does not yet establish a statistically significant truncation. As the sample of lensed star cluster candidates grows, it will enable tests of how both the slope and upper mass scale vary with redshift and environment.

The ICMF depends both on the population of star-forming clouds and on how many clusters each cloud produces, with what stellar masses. Conversion of clouds into clusters depends on gas supply, feedback and the fraction of formed stars that remains bound. Idealised cloud simulations predict how star formation responds to cloud surface density \citep{fallStellarFeedbackMolecular2010, grudicWhenFeedbackFails2018, fukushimaStarClusterFormation2020, fukushimaRadiationHydrodynamicsSimulations2021, kimModelingUVRadiation2018}. Clouds in a galaxy, however, are irregular and accrete and merge while forming stars \citep{liStarClusterFormation2017, hanImpactRadiationFeedback2022a}. Their properties also depend on the density threshold used to define them \citep{quinalhaStellarFeedbackEffects2026}. A cloud can also form several clusters, so its contribution cannot generally be described by a single efficiency applied to a fixed initial mass. Feedback can therefore change both the cloud population and the clusters each cloud produces.

 In simulations of an isolated dwarf galaxy, \citet{anderssonPresupernovaFeedbackSets2024a} found that early feedback, and ionizing radiation in particular, suppresses the most massive clusters and leaves a steep power-law ICMF, whereas SN feedback alone yields a broad and flatter cluster mass function. In the same simulations, \citet{quinalhaStellarFeedbackEffects2026} found that dense clouds ($n>100\,{\rm cm^{-3}}$) had the same mass function irrespective of feedback, while lower-density complexes ($n>10\,{\rm cm^{-3}}$) became \textit{more} top-heavy with radiative feedback. They concluded that the ICMF is shaped by feedback within each cloud and, for radiative feedback, by its effect on the temperature of the clouds. Whether these conclusions hold at cosmic dawn, when galaxies are gas-rich, compact, metal-poor and assembling through frequent mergers, has not been tested yet.

Answering this question requires simulations that follow both the cloud population and its conversion into clusters in the same galaxy, with feedback as a controlled variable. In large-scale cosmological simulations, sub-grid models such as E-MOSAICS attach clusters to star particles with a prescribed ICMF and formation efficiency \citep{pfefferEMOSAICSProjectSimulating2018, kruijssenFormationAssemblyHistory2019}, and post-processing models infer cluster populations after the fact \citep{chenCatalogueModelStar2024}, so neither predicts the ICMF. Simulations that resolve individual stars capture the conversion directly, but their computational cost limits the range of galaxies and environments that can be explored \citep{lahenGRIFFINProjectFormation2020a, hislopChallengeSimulatingStar2021}. Cosmological zooms that produce compact, feedback-free bursts build in the efficiency, either through a delay before any feedback, without radiative transfer \citep{chenFeedbackfreeStarFormation2026}, or by imposing a local star formation efficiency of unity \citep{pascaleSIEGEIVCompact2025a}.

At intermediate resolution, accreting sink particle models for star cluster formation allow the ICMF to emerge self-consistently from gas supply and stellar feedback on the scale of individual clouds \citep{liStarClusterFormation2017, liStarClusterFormation2018, brownTestingFeedbackStar2022, reina-camposStarClustersLinks2025}. We adopt such an approach in this work, and the closest precursor to our model is that described in \citet{liStarClusterFormation2017}. Their cluster particles grow inside spheres of fixed 10\,pc diameter at a rate set by a prescribed efficiency per free-fall time until their own feedback stops them from doing so. These authors find that such an approach gives rise to a Schechter-like ICMF with a low-mass slope of $\approx-1.8$ to $-2$ and a cut-off that scales with SFR.

In this paper, we present cosmological radiation-hydrodynamics zoom-in simulations of a halo of $M_{\rm vir}\sim10^{10}\,{\rm M_\odot}$ at $z\ge6$ in which star clusters, represented by sink particles, are the units of star formation and feedback. We build on the sink particle-based star formation model of \citet{kangImpactStarFormation2025, kangMitigatingOvercoolingProblem2026}, in which rapid gas accretion on to sink particles produces intense radiation that disperses star-forming gas clumps, resulting in bursty star formation. Each star cluster/sink particle accretes the amount of gas flowing into it, with no prescribed efficiency per free-fall time and no fixed formation duration. We run simulations at two different resolutions and compare runs with and without stellar radiation at the coarser resolution. We then link star clusters to their natal clouds, and compare the brightest clusters with those observed in \textit{JWST} lensed galaxies with a similar SFR. We investigate whether the brightest simulated clusters reach the observed luminosities, how stellar radiation changes the clouds and their cluster populations, and what terminates cluster growth.

Section~\ref{sec:methods} describes the simulations, the cluster sink particle model and the observational sample. Section~\ref{sec:results} presents the simulated galaxies, the ICMF and the brightest clusters, and Section~\ref{sec:discussion} traces the ICMF back to the natal clouds and the growth of sink particles, compares it to other simulations and observations and discusses the limitations of our model. Section~\ref{sec:conclusions} summarises our conclusions.

\section{Methods}
\label{sec:methods}
A detailed description of the physical models used can be found in \citet{kangImpactStarFormation2025} and \citet{kangMitigatingOvercoolingProblem2026} (referred to hereafter as \Kangfive and \Kang, respectively), the latter of which also details the initial conditions used for our simulations. We summarise the methods and initial conditions here for convenience, alongside the adaptations made to the model to capture the growth of star clusters.

{\label{subsec:simulation_suite}
Table~\ref{tab: sim sum} lists the simulation parameters. Our high-resolution run (\hires) adopts the same spatial resolution and dark-matter particle mass as the sink-particle run of \Kang, allowing a comparison with the original star formation model. The fiducial run (\fid) uses one fewer refinement level than \hires, increasing the minimum cell size by a factor of two. As the accretion radius, sink exclusion distance and density thresholds change with cell size, this comparison tests the combined resolution-dependence of the star cluster model. Finally, the run \norad has the same resolution and sink prescription as \fid but omits stellar radiation, providing a controlled feedback comparison (Section~\ref{sec:disc_resolution}).}

\begin{table}
\centering
\begin{threeparttable}
\caption{Parameters of the simulations discussed in this paper. The last column is the sink-particle simulation of \protect\Kang. $\Delta x_{\rm min}$ is the range of the smallest physical cell size over the redshift interval studied, $n_{\rm clfind}$ the density threshold of the on-the-fly clump finder, $m_{\rm dm}$ the dark-matter particle mass in the zoom region, $m_\star$ the mass of a star particle, and $r_{\rm acc}$ its accretion radius.}
\label{tab: sim sum}
\setlength{\tabcolsep}{5.8pt}
\begin{tabular}{l c c c c}
\hline \hline
 & \fid & \norad & \hires & \nc \\
\hline
$\Delta x_{\rm min}$ [pc] & 11.4--22.8 & 11.4--22.8 & 5.7--11.4 & 5.7--11.4 \\
$n_{\rm clfind}$ [cm$^{-3}$] & 25 & 25 & 100 & 100 \\
Stellar radiation & \ding{51} & \ding{55} & \ding{51} & \ding{51} \\
Cluster-sink model & \ding{51} & \ding{51} & \ding{51} & \ding{55} \\
$m_{\rm dm}$ [M$_\odot$] & 1600 & 1600 & 1600 & 1600 \\
$m_\star$ [M$_\odot$] & $10^3$ & $10^3$ & $10^3$ & $10^3$ \\
$r_{\rm acc}$ & $\Delta x_{\rm min}$ & $\Delta x_{\rm min}$ & $\Delta x_{\rm min}$ & $\Delta x_{\rm min}$ \\
Type~Ia supernovae & \ding{51} & \ding{51} & \ding{51} & \ding{55} \\
Core-collapse yields & LC18 & LC18 & LC18 & fixed \\
\hline
\end{tabular}
\end{threeparttable}
\end{table}

\subsection{Numerical framework}
\label{subsec:numerical_framework}

We perform cosmological radiation-hydrodynamics simulations with the adaptive mesh refinement code \textsc{ramses-rt} \citep{teyssierCosmologicalHydrodynamicsAdaptive2002,rosdahlRAMSESRTRadiationHydrodynamics2013,rosdahlGalaxiesThatShine2015}. The hydrodynamics equations are solved using the Harten--Lax--van Leer--Contact Riemann solver (HLLC) \citep{toroRestorationContactSurface1994} with a Courant factor of 0.8, while self-gravity is followed with a multigrid Poisson solver \citep{guilletSimpleMultigridScheme2011}. Radiative transfer is evolved using the first-moment method with the M1 closure relation \citep{levermoreRelatingEddingtonFactors1984} and a global Lax--Friedrichs flux function. To limit computational cost, we adopt a reduced speed of light, $\tilde{c}=10^{-2}c$, well above the propagation speed of ionization fronts in the star-forming regions of interest.

Radiation is discretised into six photon groups with energy boundaries approximately given by $5.6$, $11.2$, $13.6$, $15.2$, $24.6$, and $54.4\,{\rm eV}$, with the highest-energy bin extending to infinity. These groups are chosen to capture molecular hydrogen dissociation, hydrogen and helium ionization, photoheating, and direct radiation pressure. The radiation field is coupled to a non-equilibrium primordial chemistry network for $\mathrm{H_2}$, \ion{H}{i}, \ion{H}{ii}, \ion{He}{i}, \ion{He}{ii}, \ion{He}{iii}, and free electrons \citep{rosdahlRAMSESRTRadiationHydrodynamics2013,katzInterpretingALMAObservations2017}. We also include a spatially uniform, redshift-dependent ultraviolet background \citep{haardtRadiativeTransferClumpy2012} when computing photoionization and photoheating rates. It is damped before reionization, by a factor $10^{-20(z/8.5)^6}$, so that it is negligible until $z\approx7$ and it is attenuated in each cell by the self-shielding factor $\exp(-n_{\rm H}/0.01\,{\rm cm^{-3}})$. This correction applies to the background only. Radiation from the stars in the simulation is transported explicitly by the radiative-transfer solver, so its absorption is followed cell by cell.

Gas cooling includes both primordial and metal-line contributions. At temperatures $T \gtrsim 10^4\,{\rm K}$, metal-line cooling is computed from tabulated rates as a function of density, temperature, and metallicity using \textsc{cloudy}-based tables \citep{ferlandCLOUDY90Numerical1998}. At lower temperatures, fine-structure cooling follows \citet{rosenGlobalModelsInterstellar1995}. We also include dust following a metallicity-dependent scaling for gas with $T<10^5\,{\rm K}$ based on \citet{remy-ruyerGastodustMassRatios2014}.

\subsection{Cosmological initial conditions and refinement}
\label{subsec:initial_conditions}
The simulations use the same cosmological zoom-in initial conditions as the reference model of \Kang. The initial conditions were generated with \textsc{genetIC} \citep{stopyraGenetICNewInitial2021}, adopting Planck-compatible cosmological parameters \citepalias{collaborationPlanck2018Results2020}, namely $\Omega_{\rm m}=0.3158$, $\Omega_{\Lambda}=0.6842$, $\Omega_{\rm b}=0.045$, and $H_0=67.32\,{\rm km\,s^{-1}\,Mpc^{-1}}$. A dark-matter-only parent simulation was first run in a $(30\,{\rm cMpc})^3$ volume with $256^3$ particles. From this volume, a halo with $M_{\rm vir}\sim10^{10}\,{\rm M_\odot}$ at $z=6$ was selected for resimulation. The high-resolution region extends to three virial radii around the target halo, within which the dark matter particle mass is $m_{\rm dm}\simeq1600\,{\rm M_\odot}$. The bulk velocity of the zoom region was reduced using the method of \citet{pontzenEDGENewApproach2021}, giving a halo velocity of approximately $16\,{\rm km\,s^{-1}}$ at $z=6$ and thereby limiting advection-related errors in the grid calculations.

The simulation domain is initially divided into $256^3$ cells and is adaptively refined within the zoom region. A cell is refined if the total mass in dark matter and baryons, divided by the cosmic baryon fraction $\Omega_{\rm b}/\Omega_{\rm m}$, exceeds $8m_{\rm dm}\simeq1.24\times10^4\,{\rm M_\odot}$. We also enforce refinement when the thermal Jeans length,
\begin{equation}
    \lambda_{\rm J} = \sqrt{\frac{\pi c_{\rm s}^2}{G\rho_{\rm gas}}},
\end{equation}
is resolved by fewer than 16 cells, where $c_{\rm s}$ and $\rho_{\rm gas}$ are the isothermal sound speed and gas density of the cell. All cells inside the sink particle accretion radius are kept at the maximum refinement level. Additional refinement levels are released at $z=19$ and $z=9$, corresponding to $a=0.05$ and $a=0.1$, to keep the physical resolution approximately constant. As a result, in \fid, the minimum cell size remains in the range $\Delta x_{\rm min}=11.4$--$22.8\,{\rm pc}$, and $\Delta x_{\rm min}=5.7$--$11.4\,{\rm pc}$ in the \hires simulation, over the redshift interval studied.

Haloes are identified using \textsc{rockstar-galaxies} \citep{behrooziROCKSTARPhasespaceTemporal2013} applied to the dark matter particle distribution. We define the virial radius, $R_{\rm vir}$, as the radius enclosing the mean density $\Delta_{\rm crit}\rho_{\rm crit}(z)$ of \citet{bryanStatisticalPropertiesXRay1998}, and the virial mass $M_{\rm vir}$ as the mass inside it. At early times, the halo centre is taken to be the dark matter density peak. Once the galaxy contains more than $10^5\,{\rm M_\odot}$ in stars (including cluster sink particles), we instead define the centre using a shrinking sphere method on the stellar mass distribution \citep{powerInnerStructureLambdaCDM2003}, and recompute $R_{\rm vir}$ about this centre.

\subsection{Star formation and cluster sink particles}
\label{subsec:sink_star}

We employ the Pop~III prescription of \citet{kimmFeedbackregulatedStarFormation2017}. Pop~III stars form in maximally refined cells where the gas flow is convergent and the gas metallicity $Z_{\rm gas}\leq10^{-6}Z_\odot$ (with $Z_\odot=0.0134$ \citep{asplundChemicalCompositionSun2009a}), once the local turbulent Jeans length falls below the minimum cell size. The masses of individual Pop-III stars are sampled between 10 and $10^3\,{\rm M_\odot}$ from a Salpeter-like IMF with an exponential suppression below $100\,{\rm M_\odot}$ \citep{salpeterLuminosityFunctionStellar1955,wiseBirthGalaxyPrimordial2012a}.

\subsubsection{Seeding, accretion and stellar populations}

Our Pop~II star formation model is based on the sink particle prescription of \Kangfive and \Kang, but modified so that, depending on their mass, sink particles represent either individual forming star clusters or temporary reservoirs from which collisionless star particles are spawned. Candidate star-forming gas clumps are identified on the fly with the \textsc{phew} algorithm \citep{bleulerMoreRealisticSink2014, bleulerPHEWParallelSegmentation2015}, which segments the gas density field into isodensity contours above a threshold $n_{\rm clfind}$. We use $n_{\rm clfind}=25\,{\rm cm^{-3}}$ in \fid and \norad and $100\,{\rm cm^{-3}}$ in \hires, following the $\rho_{\rm th}\propto\Delta x_{\rm min}^{-2}$ scaling of equation~(\ref{eq: LP condition}) below. A gas density peak is kept as a separate clump only if its density exceeds that of the saddle point connecting it to a neighbouring clump by at least a factor of 2 (the relevance threshold), and is merged with it otherwise.
The clouds analysed in Section~\ref{sec:discussion} are identified separately in post-processing, without performing this saddle point density splitting step (Section~\ref{sec:clouds}). A sink particle is seeded at the density peak of a clump with an initial mass of $1\,{\rm M_\odot}$ when all of the following conditions hold: the peak cell is maximally refined, the gas is enriched above the Pop~III metallicity threshold, the local gas velocity field is converging, the clump is virialised, and no other sink particle lies within a distance of twice the sink particle accretion radius, $2r_{\rm acc}$, from the peak ($r_{\rm acc} =\Delta x_{\rm min}$ in all our runs). In addition, the peak density must exceed
\begin{equation}
    \rho_{\rm th} =
    \frac{8.86c_{\rm s}^2}{\pi G\Delta x_{\rm min}^2},
    \label{eq: LP condition}
\end{equation}
which corresponds to the density $0.5\Delta x_{\rm min}$ away from the centre of a collapsing isothermal sphere \citep{larsonNumericalCalculationsDynamics1969,penstonDynamicsSelfgravitatingGaseous1969,gongImplementationSinkParticles2013}.

Once formed, sink particles grow by accreting gas according to a flux-based scheme. The accretion rate on to each sink particle is calculated from the net gas flux through the surface of a spherical accretion zone of radius $r_{\rm acc}$,
\begin{equation}
    \dot{M}_{\rm acc}
    =
    \max\left[0,\;
    -\oint
    \rho_{\rm gas}
    \left(
    \boldsymbol{v}_{\rm gas}
    -
    \boldsymbol{v}_{\rm sink}
    \right)
    \cdot
    \hat{\boldsymbol{n}}\,{\rm d}S
    \right],
\end{equation}
where $\boldsymbol{v}_{\rm gas}$ is the gas velocity, $\boldsymbol{v}_{\rm sink}$ is the sink particle velocity, and $\hat{\boldsymbol{n}}$ is the outward unit vector normal to the accretion surface, $S$. We use the flux-based rate rather than a Bondi--Hoyle--Lyttleton estimate because the latter would require an ambient density and velocity at infinity, which are not well defined in a turbulent, inhomogeneous ISM. To avoid numerical instabilities, the timestep is limited so that a sink particle cannot double its mass in a single fine step and no cell can lose more than 75~per~cent of its gas mass \citep{bleulerMoreRealisticSink2014}. Pairs of sinks whose accretion regions overlap and that are gravitationally bound are merged, with total mass and momentum conserved.

In the original implementation of \Kangfive and \Kang, a sink particle acts as a reservoir. Whenever it accretes a mass equivalent to a stellar particle ($1000\,{\rm M_\odot}$ in \Kang and this study, $500\,{\rm M_\odot}$ in \Kangfive) this mass is converted into a collisionless star particle that sources radiative and supernova feedback. A sink particle is not destroyed after the formation of the first $1000\,{\rm M_\odot}$ star particle, but can continue to accrete and form other star particles of the same mass. We adapt this procedure to model (unresolved) star cluster formation. In our implementation, all the star particles that a sink particle forms remain attached to their parent sink particle rather than becoming freely orbiting collisionless particles. The sink particle therefore represents a forming star cluster that carries one or more embedded stellar populations, each with a unique age and metallicity. We retain the term 'star particle' for these attached $10^3\,{\rm M_\odot}$ populations even though they are technically part of a larger star cluster. Radiation, supernovae, and chemical enrichment from these star particles are injected at the position of the parent sink particle. If two sink particles merge, the star particles associated with both progenitors are assigned to the remnant sink particle. As a result, each sink particle constitutes an unresolved, sub-grid star cluster. We do not attempt to follow the internal structure or bound mass of the cluster here as this will be explored in a companion paper. We refer to sink particles when describing their formation and growth, and to clusters when comparing them with observed star clusters or with star clusters in other simulations.

\subsubsection{End of formation and retention}

We declare that a sink particle's star formation episode has ended after its local gas supply has remained dynamically insignificant over a long enough period of time. To estimate the length of this period, we compute the mean density inside the accretion region,
\begin{equation}
    \rho_{\rm cl}
    =
    \frac{
    M_{\rm sink}
    +
    M_{\rm gas}(<r_{\rm acc})
    }
    {
    \frac{4}{3}\pi r_{\rm acc}^3
    },
\end{equation}
and the corresponding free-fall time,
\begin{equation}
    t_{\rm ff}
    =
    \sqrt{
    \frac{3\pi}{32G\rho_{\rm cl}}
    }.
\end{equation}
We then monitor the gas mass we expect the sink particle to accrete within three local free-fall times, $3\dot{M}t_{\rm ff}$, where $\dot M$ is the net mass flux on to the sink particle. If this mass stays below an individual stellar particle mass ($10^3\,{\rm M_\odot}$) for at least $1\,{\rm Myr}$, the cluster is considered formed and any further gas accretion on to the sink particle is disabled. The duration of this lull is a numerical choice. It determines when the formation phase of a sink is declared over, whereas the end of accretion itself is usually set by the reversal of the gas inflow (Section~\ref{sec:shutdown}). For all but 0.2~per~cent of the sink particles, the accretion rate is already exactly zero throughout this Myr. Accretion therefore ends before the formation phase is declared over, with its termination set by the gas flow rather than this numerical criterion. Once this happens, gas accretion on to the sink particle is prevented from restarting at a later time. The one exception is mergers. If either of two merging sink particles is still accreting, the remnant can accrete as well, so a star cluster whose formation had ended can resume growing if it merges with a forming one. Finally, any remaining non-stellar sink particle mass is converted into a final low-mass star particle when cluster formation ends, and this particle is prevented from producing supernovae if its mass is too small to allow a consistent sampling of the IMF.

After the initial formation phase, if the associated stellar mass satisfies $M_{\rm cl}\geq10^4\,{\rm M_\odot}$, the sink particle is retained as a long-lived cluster particle, with its star particles attached. If $M_{\rm cl}<10^4\,{\rm M_\odot}$, the sink particle is destroyed and its star particles are redistributed as independent collisionless particles, with positions drawn from a Plummer sphere of scale radius $\Delta x_{\rm min}$ centred on the former sink particle, conserving total mass and momentum. This threshold is a bookkeeping choice that limits the number of long-lived sink particles. It does not remove low-mass objects from the recorded birth catalogue, because every sink particle is followed until cluster formation ends and all birth masses are recorded (Section~\ref{sec:icmf}).

\subsection{Stellar feedback}
\label{subsec:fbk}

The stellar feedback model follows \Kang, with the same parameter choices unless otherwise stated. We include radiation from young stellar populations, core-collapse supernovae from Pop~II stars, Type~Ia supernovae (not included in \Kang), and mass-dependent explosions of Pop~III stars. No additional feedback channel is introduced specifically for star clusters. The main change is geometric and temporal. In the cluster-sink model, each sink particle hosts the star particles it forms or acquires via inter-sink particle mergers, and their feedback is injected at the parent sink position. Concentrating the feedback at one position couples time-resolved sink growth directly to the radiation, mass, metals, momentum and energy returned by its attached stellar populations.

Radiative feedback follows \citet{kimmFeedbackregulatedStarFormation2017}. Each star particle emits photons with age- and metallicity-dependent luminosities in the six radiation groups described above. For Pop~II star populations, we use \textsc{bpass} v2.2.1 stellar population models \citep{stanwayStellarPopulationEffects2016,stanwayReevaluatingOldStellar2018}, assuming a \citet{kroupaVariationInitialMass2001} IMF with lower and upper stellar mass limits of $0.1$ and $300\,{\rm M_\odot}$, respectively. Pop~III stellar lifetimes and photon yields are taken from \citet{schaererPropertiesMassivePopulation2002}. Ionizing photons photoionize \ion{H}{i}, \ion{He}{i}, \ion{He}{ii}, and $\mathrm{H_2}$ according to the relevant cross sections, while Lyman--Werner photons dissociate $\mathrm{H_2}$. Absorbed photons heat the gas through excess photoionization energy and transfer momentum to it through direct radiation pressure.

Core-collapse supernovae are coupled to the gas using the mechanical feedback model of \citet{kimmEscapeFractionIonizing2014,kimmSimulatingStarFormation2015}. All stars with initial masses above $8\,{\rm M_\odot}$ are assumed to explode, corresponding to about one supernova per $100\,{\rm M_\odot}$ of stars formed for our adopted IMF. Each explosion injects $10^{51}\,{\rm erg}$ of energy and deposits radial momentum into neighbouring gas cells. The time delays for the supernovae of each star particle are drawn independently from a fit to the cumulative supernova number of a \textsc{starburst99} population \citep{kimmSimulatingStarFormation2015}. The delay model allows explosions between approximately 3.4 and 44\,Myr after star-particle formation. We use its minimum allowed delay of $\sim3.4\,{\rm Myr}$ after $t_{\rm first}$ as a common supernova reference in all three simulations. If the Sedov--Taylor phase of the explosion is unresolved, the injected momentum is set to a terminal value calibrated on resolved supernova remnant calculations, with the terminal momentum depending on the local density and metallicity.

Unlike \Kang, Pop~II core-collapse supernovae return individual chemical species (H, He, C, N, O, Ne, Mg, Si, S, Ca and Fe) with the mass- and metallicity-dependent yields of the rotating $150\,{\rm km\,s^{-1}}$ models of \citet{limongiPresupernovaEvolutionExplosive2018}.

Type~Ia supernovae are included as a delayed enrichment channel that couples only thermally to the gas. Delay times follow a $t^{-1}$ distribution between 50\,Myr and 13.7\,Gyr \citep{maozDelaytimeDistributionTypeIa2012}, normalized to $1.3\times10^{-3}$ events per ${\rm M_\odot}$ of stars formed, and each event injects $10^{51}\,{\rm erg}$ of thermal energy and $1.4\,{\rm M_\odot}$ of ejecta enriched in metals with the N100 yields of \citet{seitenzahlThreedimensionalDelayeddetonationModels2013}.

Pop~III stars with masses of 11--20, 20--40 and 140--260\,${\rm M_\odot}$ explode as core-collapse supernovae, hypernovae and pair-instability supernovae respectively, with the same energies and yields as in \Kang \citep{nomotoNucleosynthesisYieldsCorecollapse2006, hegerNucleosyntheticSignaturePopulation2002}. Those with other masses collapse directly into black holes.

\subsection{Cluster photometry and the observational sample}
\label{subsec:light}
\label{subsec:obs}
We compute the luminosity of each simulated cluster from the stellar populations it carries, taking into account their ages and metallicities, with the same \textsc{bpass} v2.2.1 models that drive the on-the-fly radiative transfer in the simulations (Section~\ref{subsec:fbk}). A cluster's light includes every star it holds, including those brought in by mergers with other clusters. We quote intrinsic, stellar-only rest-frame AB magnitudes \citep{okeSecondaryStandardStars1983} at 1500\,\AA, $M_{\rm UV}$, without nebular continuum, emission lines or dust attenuation. For young clusters, the contribution from the nebular continuum is likely non-negligible, and we discuss its potential impact on our comparison to observed clusters in Section~\ref{sec:disc_obs}.

We compare with the brightest compact sources in ten strongly lensed systems observed by \textit{JWST} between $z=4.9$ and 11.5, listed in Table~\ref{tab:obs}. Eight come from the compilation of \citet{pfefferComparingEMOSAICSPredictions2025}, namely RCS0224, the Sunrise arc, the Cosmic Grapes, the three MACS0416 subsystems, the Firefly Sparkle and the Cosmic Gems arc, and we add the Cosmic Spear \citep{abdurroufSpatiallyResolvedPhysical2025} and the Misty Moons \citep{nakaneVENUSStronglyLensed2025}. We consider observed systems for which the brightest source has rest-frame UV photometry and the host galaxy SFR is averaged over a stated time interval, either published or derived in this work from published recombination-line fluxes. Appendix~\ref{app:obs} lists the sample and describes how we reduce each observed magnitude to an intrinsic rest-frame value at 1500\,\AA, as well as how the host galaxy SFRs are estimated.

\section{Results}
\label{sec:results}

%==============================================================================
\subsection{The simulated galaxies}
\label{sec:galaxies}

We first establish the galactic environment in which the clusters form and examine whether the cluster-sink prescription changes the integrated properties and evolution of the galaxy. We compare \hires, \fid and \norad with \nc, which shares our initial conditions and the resolution of \hires but uses the original non-cluster sink model (Section~\ref{subsec:simulation_suite}). Table~\ref{tab: z=6} summarises the characteristics of the main galaxy and its cluster population at $z=6$. We define the central galaxy by a spherical aperture of radius $0.2\,R_{\rm vir}$, as in \Kang, centred as described in Section~\ref{subsec:initial_conditions}.

\begin{table}
\centering
\begin{threeparttable}
\caption{The main galaxy and its clusters at $z=6$. Virial and stellar masses are calculated inside $R_{\rm vir}$, and metallicities, SFR, UV magnitude and gas properties, including $M_{\rm gas}$ and the gas fraction, inside $0.2\,R_{\rm vir}$. The gas fraction is calculated using the stellar mass inside the same aperture. Metallicities are mass-weighted, in units of $Z_\odot=0.0134$. $M_{\rm UV}$ is intrinsic (stellar light only). $\mathrm{SFR}_{100}$ is averaged over the final 100~Myr ending at $z=6$. $r_{1/2}$ and $\Sigma_{\rm gas}=M_{\rm gas}/(2\pi r_{1/2}^2)$ refer to the gas half-mass radius and surface density respectively. $N_{\rm cl}$ is the number of analysed cluster birth episodes completed by $z=6$ (Section~\ref{subsec:selection}). This count is not defined for \nc, whose sink particles do not represent clusters.}
\label{tab: z=6}
\begin{tabular}{l c c c c}
\hline \hline
 & \hires & \fid & \norad & \nc \\
\hline
$M_{\rm vir}$ [$10^{10}\,\mathrm{M_\odot}$] & 1.17 & 1.14 & 1.18 & 1.18 \\
$M_\star$ [$10^8\,\mathrm{M_\odot}$] & 2.23 & 2.04 & 3.78 & 2.15 \\
$Z_{\rm gas}$ [$Z_\odot$] & 0.44 & 0.36 & 0.64 & 0.20 \\
$Z_\star$ [$Z_\odot$] & 0.25 & 0.19 & 0.51 & 0.17 \\
$\mathrm{SFR}_{100}$ [$\mathrm{M_\odot\,yr^{-1}}$] & 1.15 & 0.80 & 2.80 & 0.86 \\
$M_{\rm UV}$ [AB mag] & $-18.46$ & $-18.64$ & $-19.60$ & $-18.53$ \\
$M_{\rm gas}$ [$10^8\,\mathrm{M_\odot}$] & 2.87 & 2.55 & 3.40 & 3.4 \\
$M_{\rm gas}/(M_{\rm gas}+M_\star)$ & 0.59 & 0.58 & 0.50 & 0.60 \\
$r_{1/2}$ [kpc] & 1.12 & 0.98 & 0.92 & 1.14 \\
$\Sigma_{\rm gas}$ [$\mathrm{M_\odot\,pc^{-2}}$] & 36 & 42 & 63 & 41 \\
$N_{\rm cl}$ & 1469 & 385 & 890 & -- \\
\hline
\end{tabular}
\end{threeparttable}
\end{table}

\begin{figure}
    \includegraphics[width=\linewidth]{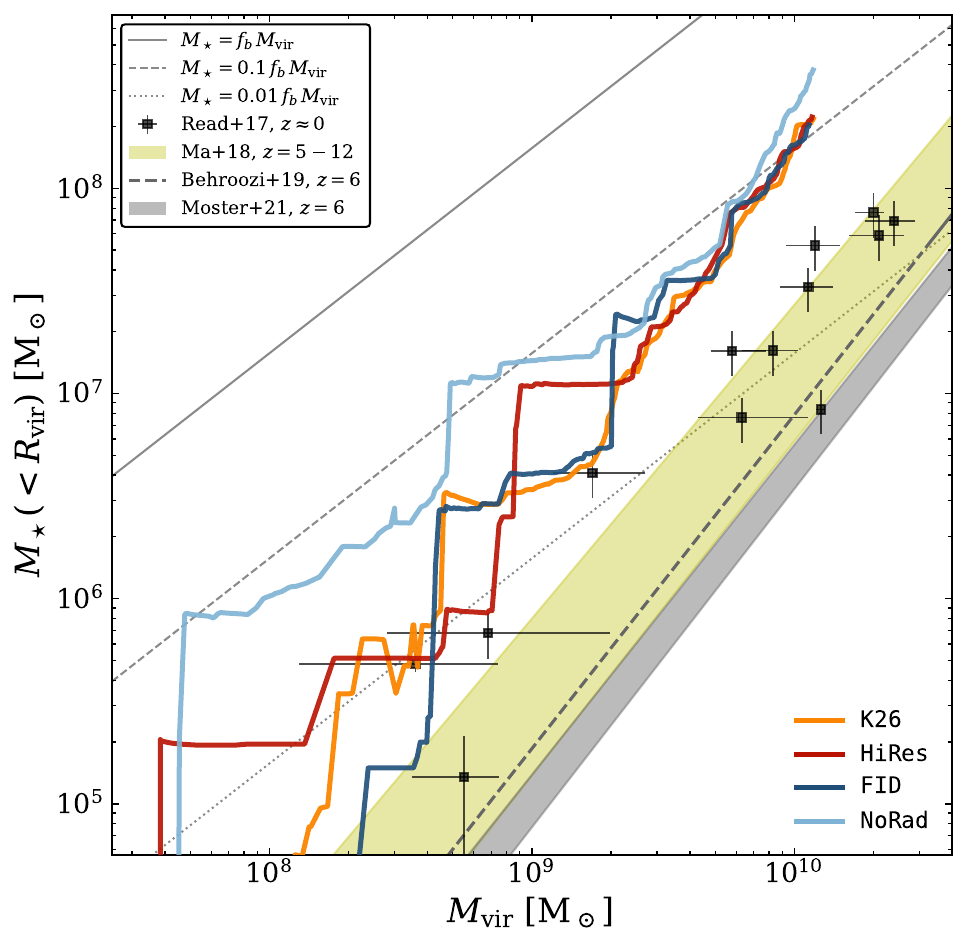}
    \vspace{-5mm}
    \caption{Stellar mass against virial mass of the main halo along its assembly history ($z\approx16$--6), for \hires, \fid, \norad and the non-cluster sink particle run of \protect\Kang (\nc, thick solid lines, identified in the legend). Diagonal lines mark $M_\star=f_{\rm b}M_{\rm vir}$ (solid), $0.1\,f_{\rm b}M_{\rm vir}$ (dashed) and $0.01\,f_{\rm b}M_{\rm vir}$ (dotted), where $f_{\rm b}=\Omega_{\rm b}/\Omega_{\rm m}$ is the cosmic baryon fraction. Also shown are $z\approx0$ dwarfs (squares, \citealt{readStellarMassHalo2017}), the relation from the FIRE-2 simulations at $z=5$--12 (light band, \citealt{maSimulatingGalaxiesReionization2018}) and the $z=6$ relations of UniverseMachine (thick dashed, \citealt{behrooziUniverseMachineCorrelationGalaxy2019}) and GalaxyNet empirical models (dark band, \citealt{mosterGalaxyNetConnectingGalaxies2021}).}
    \alttext{Logarithmic plot of stellar mass versus halo mass. HiRes, FID and K26 follow similar growth tracks, while NoRad reaches a higher stellar mass. At large halo masses, all four tracks lie above the FIRE-2 and empirical relations. Diagonal lines show reference fractions of the cosmic baryon budget.}
    \label{fig:smhm}
    \vspace{-3mm}
\end{figure}

\hires and \fid track the stellar mass--halo mass relation of \nc closely (Fig.~\ref{fig:smhm}), ending within 10~per~cent of its stellar mass by $z=6$. Despite the changes to the cluster, yield and Type~Ia models, the stellar mass growth of the galaxy is therefore essentially unchanged, as it is when the spatial resolution of the simulation is increased by a factor 2. Without stellar radiation the galaxy forms more stars at almost every halo mass and ends up with nearly twice the stellar mass of \hires and \fid. All runs lie above the FIRE-2 relation \citep{maSimulatingGalaxiesReionization2018} and the $z=6$ relations of empirical models \citep{behrooziUniverseMachineCorrelationGalaxy2019,mosterGalaxyNetConnectingGalaxies2021}, by about an order of magnitude in stellar mass at fixed halo mass. This excess indicates a less efficient regulation of galaxy-wide star formation in comparison to these models. Additional feedback channels, including cosmic rays, are possible contributors \citep{rodriguezmonteroImpactCosmicRays2024}, but their effects are not isolated by our experiment.

\begin{figure}
    \includegraphics[width=\linewidth]{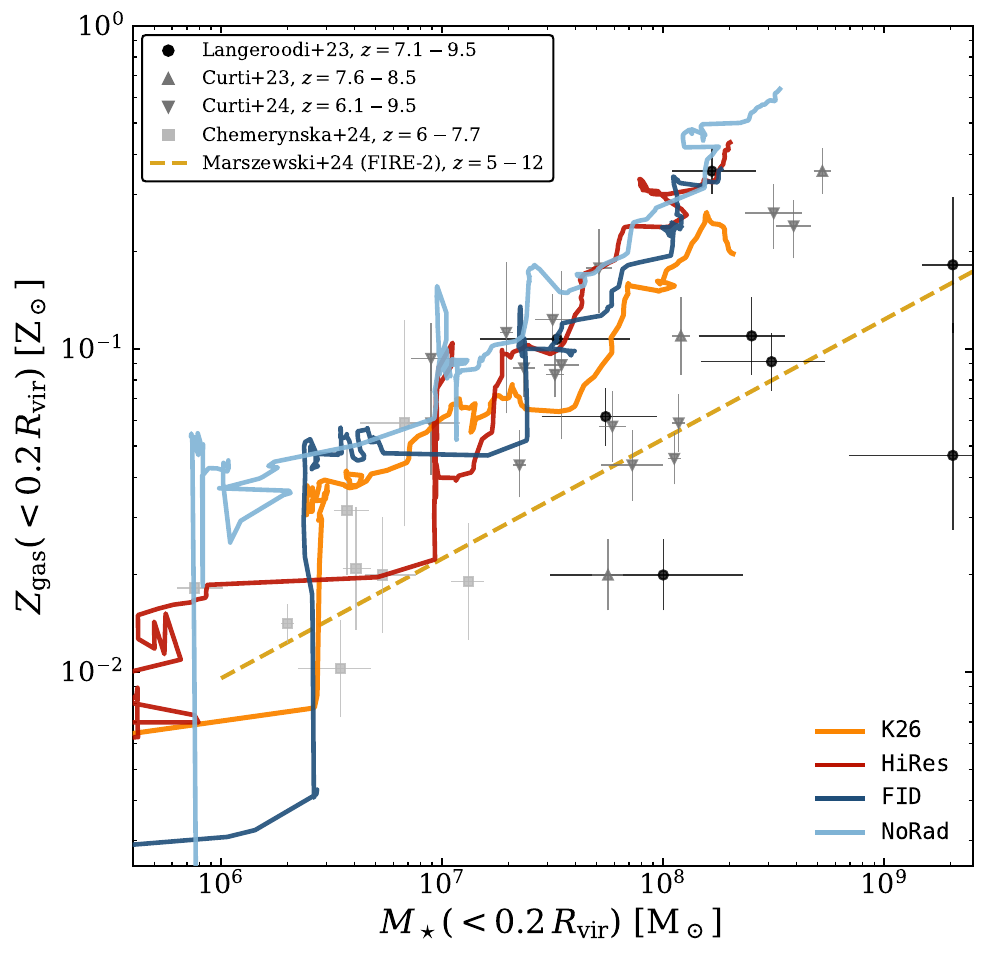}
    \vspace{-5mm}
    \caption{Mass-weighted gas metallicity against stellar mass, both inside $0.2\,R_{\rm vir}$, for \hires, \fid, \norad and \nc (solid lines, identified in the legend). Observed oxygen abundances are converted to $Z/Z_\odot$ assuming solar abundance ratios \citep{asplundChemicalCompositionSun2009a}. \textit{JWST} measurements are from \citet[][circles]{langeroodiEvolutionMassMetallicityRelation2023}, \citet[][upward triangles]{curtiChemicalEnrichmentEarly2023}, \citet[][downward triangles]{curtiJADESInsightsLowmass2024} and \citet[][squares]{chemerynskaExtremeLowmassEnd2024}, all at $z\approx6$--9.5. The dashed line shows the relation from the FIRE-2 simulations at $z=5$--12 \citep{marszewskiHighRedshiftGasPhaseMassMetallicity2024}.}
    \alttext{Logarithmic plot of gas metallicity versus stellar mass, comparing four simulated growth tracks with observed galaxies and the FIRE-2 relation. Metallicity generally rises with stellar mass. HiRes and FID are more enriched than K26, while NoRad reaches the highest metallicity. The tracks overlap the upper part of the observed distribution.}
    \label{fig:mzr}
    \vspace{-3mm}
\end{figure}

\hires and \fid are more metal-rich than \nc at fixed stellar mass, by a factor of $\approx1.3$--1.5 (Fig.~\ref{fig:mzr}). The larger metal yield of our enrichment model can account for much of this offset. We adopt the yields of \citet{limongiPresupernovaEvolutionExplosive2018} for rotating massive stars, whereas \nc use a fixed metal yield \citep{kangMitigatingOvercoolingProblem2026}. Our runs thus return about 1.5 times more new metals per unit stellar mass formed. Type~Ia supernovae, absent from \nc, supply only 2~per~cent of the metals returned by $z=6$ and are unlikely to explain most of the offset. \norad is more enriched still, because it forms more stars at fixed halo mass. Above $M_\star\approx10^7\,{\rm M_\odot}$ all runs, including \nc, systematically lie within the upper envelope of \textit{JWST} measurements. We measure the mass-weighted total gas metallicity. Observations give nebular oxygen abundances, converted here assuming solar abundance ratios.

\begin{figure}
    \includegraphics[width=\linewidth]{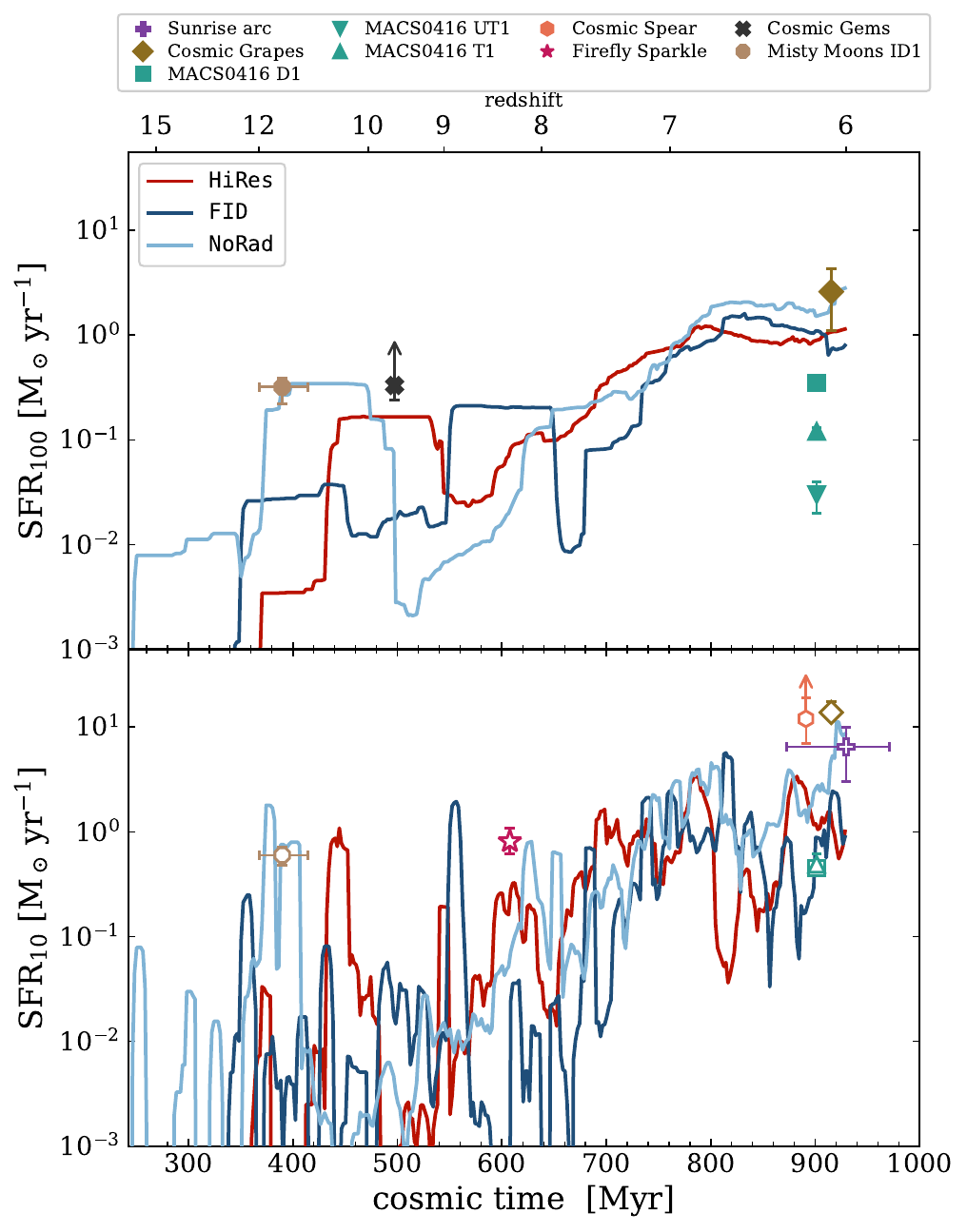}
    \vspace{-5mm}
    \caption{Star formation history of the main galaxy, averaged over 100 (top) and 10~Myr (bottom), and calculated from the accretion on to sink particles located within $0.2 R_{\rm vir}$, for \hires, \fid and \norad (legend in the top panel). Symbols show the lensed \textit{JWST} host galaxies of Table~\ref{tab:obs} at their own redshift (RCS0224, at $z=4.88$, lies beyond the time axis), with the published SFR averaged over about 100~Myr in the top panel (filled, as in Fig.~\ref{fig:brightest}) and about 10~Myr in the bottom panel (open), and arrows marking lower limits.}
    \alttext{Two panels show star formation rate versus cosmic time, averaged over one hundred million years above and ten million years below. All three simulations show bursts and intervening lulls; shorter averaging reveals much sharper fluctuations. Rates generally rise towards later times. Symbols and lower-limit arrows mark observed lensed host galaxies.}
    \label{fig:sfr}
    \vspace{-3mm}
\end{figure}

We estimate SFRs from gas accretion on to sink particles inside the central aperture. Star formation is bursty in all three runs (Fig.~\ref{fig:sfr}), as commonly found in simulations of low-mass galaxies with strong stellar feedback, where SFRs can easily vary by an order of magnitude over a few Myr \citep[e.g.][]{sparreStarburstsFIREObservational2017}. Averaged over 10~Myr, the SFR typically peaks at three to five times its 100~Myr peak value. Such bursts usually coincide with pericentric passages of satellite galaxies and major mergers, after which there are often periods where star formation ceases entirely. The 10 and 100~Myr averages at a given instant can therefore differ widely, which matters when comparing with observations whose SFR tracers probe different time-scales (Sections~\ref{subsec:obs} and \ref{sec:brightest}). Note that the peak $\mathrm{SFR}_{100}$ of our main galaxy lies within the range of the lensed \textit{JWST} host galaxies at $z\approx6$, although these span a range of masses and environments.

The main galaxy is gas-rich and compact (Table~\ref{tab: z=6}). A few $10^8\,{\rm M_\odot}$ of gas, i.e. half or more of the baryonic mass inside $0.2\,R_{\rm vir}$, lies within a half-mass radius of $\approx1$\,kpc, and our star clusters form from metal-poor gas, at a median $\approx0.1\,Z_\odot$. We compare these properties with those of isolated-galaxy simulations in Section~\ref{sec:disc_sims}.

%==============================================================================
\subsection{The initial cluster mass function}
\label{sec:icmf}
\label{subsec:samples}
\label{subsec:icmffit}

\label{subsec:selection}
Each sink particle represents an unresolved star cluster in formation. We define each sink particle's \emph{birth mass}, $M_{\star,\rm birth}$, as the total stellar mass formed during the first accretion episode. Mass gained after that episode, through mergers or renewed accretion, is excluded. For the ICMF and subsequent cluster-formation analyses, we select completed first accretion episodes without mergers, requiring the first star particle to form within the central aperture, a measured end of accretion and at least one star particle formed before that time. This gives 1469, 385 and 890 clusters in \hires, \fid and \norad, respectively. The sample includes low-mass sink particles that are subsequently converted into field stars (Section~\ref{subsec:sink_star}). Birth masses measure stars formed, with no correction for unresolved gas-expulsion or dynamical losses, or for the fraction that remains bound. The brightest-cluster comparison instead selects clusters at the time of observation (Section~\ref{sec:brightest}).

We fit the ICMF above a lower mass threshold $M_{\rm min}$ with both a power law, ${\rm d}N/{\rm d}M\propto M^{\beta_{\rm pl}}$, and an exponentially truncated power law, ${\rm d}N/{\rm d}M\propto M^{\beta_{\rm tr}}\exp(-M/M_{\rm c})$ \citep{schechterAnalyticExpressionLuminosity1976a}, using the unbinned likelihood of the clusters' individual masses \citep{johnsonPanchromaticHubbleAndromeda2017,messaYoungStarCluster2018} and sampling the posterior with \textsc{emcee} \citep{foreman-mackeyEmceeMCMCHammer2013} under priors flat in the slopes and in $\log_{10}M_{\rm c}$, with $10^4<M_{\rm c}<10^{10}\,{\rm M_\odot}$. We quote posterior medians with 16th--84th percentiles. The fitted slope and cut-off depend on $M_{\rm min}$, so we scan $M_{\rm min}$ discretely in each run and pick the lowest value whose Kolmogorov--Smirnov distance is within 3~per~cent of the global minimum for the fitted truncated power law. We then test the goodness of fit of both laws with synthetic samples drawn from the fitted ICMFs above $M_{\rm min}$ \citep{clausetPowerlawDistributionsEmpirical2009}, and compare one run to another using their own $M_{\rm min}$ values (see Appendix~\ref{app:icmf} for details).

\begin{figure*}
    \includegraphics[width=\textwidth]{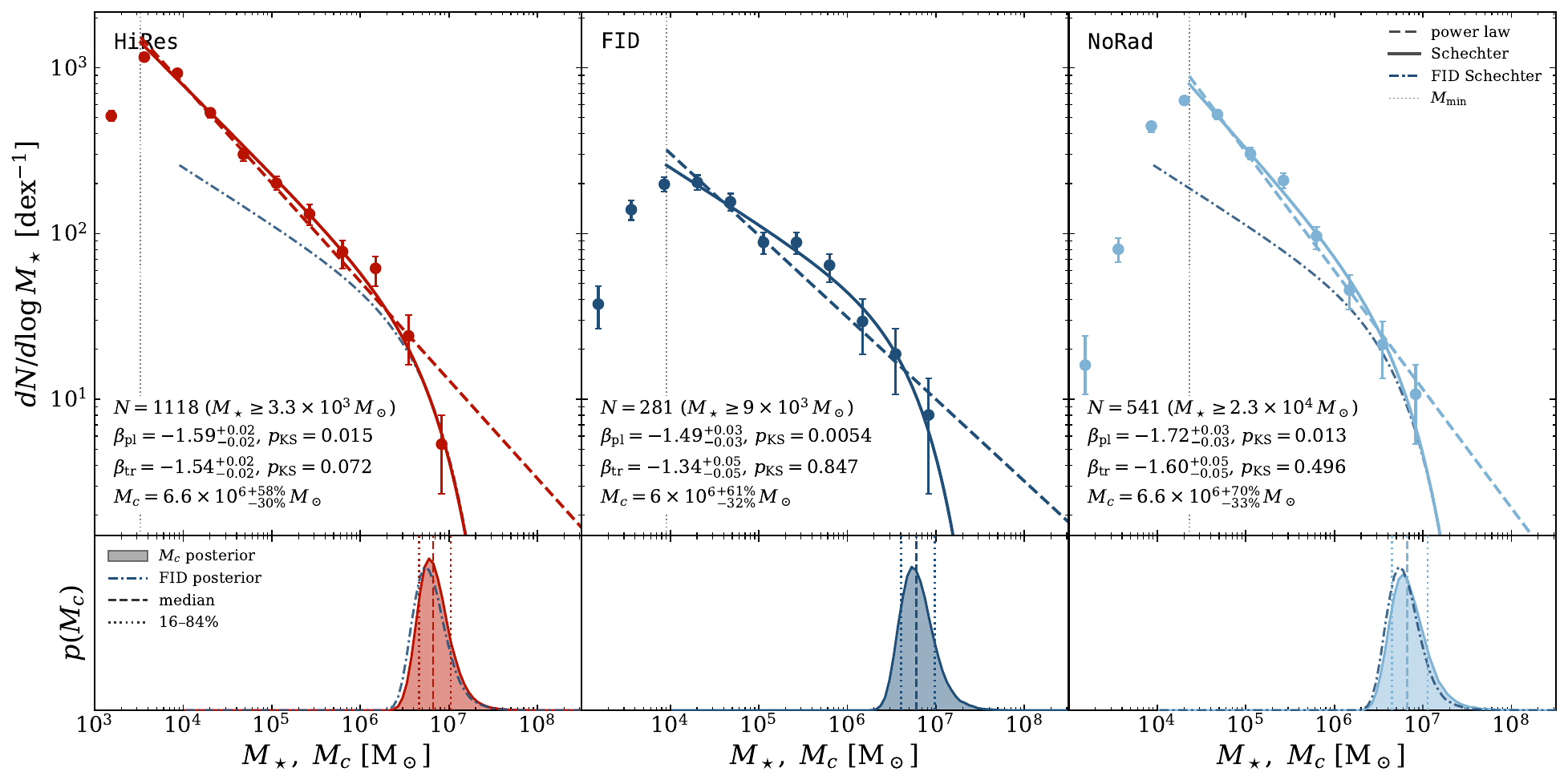}
    \vspace{-5mm}
    \caption{Initial cluster mass function (ICMF), the total number of star clusters per dex of birth mass, ${\rm d}N/{\rm d}\log M_\star$, born in the main galaxy of the \hires (left), \fid (centre) and \norad (right) simulations, with 16th--84th percentile bootstrap error bars. Dashed and solid lines are the posterior-median power law and truncated (Schechter) law fitted above each run's minimum birth mass $M_{\rm min}$ (vertical dotted line), chosen with the Kolmogorov--Smirnov scan of Appendix~\ref{app:icmf}. The dash-dotted curve in the left and right panels repeats the truncated fit of \fid. Each panel lists the number of clusters with masses above $M_{\rm min}$, the fit parameters and the Kolmogorov--Smirnov goodness-of-fit $p$-value of each fitted law, calculated by repeating the lower mass limit selection and fitting procedure of Appendix~\ref{app:icmf}. Bottom panels show the marginal posterior of the cut-off mass from each Schechter fit, with its median (dashed) and 16th and 84th percentiles (dotted), and the posterior of \fid repeated unfilled in the left and right panels (dash-dotted). }
    \alttext{Three columns compare initial cluster mass functions in HiRes, FID and NoRad. Counts fall with increasing birth mass and steepen at the high-mass end. FID has a shallower fitted slope than the other runs. The lower panels show strongly overlapping posterior distributions for the Schechter cut-off mass; vertical lines in the upper panels mark the lower fitting limits.}
    \label{fig:icmf}
    \vspace{-3mm}
\end{figure*}

The Schechter fits yield $M_{\rm c}=6.6^{+3.8}_{-2.0}$, $6.0^{+3.7}_{-1.9}$ and $6.6^{+4.6}_{-2.2}\times10^6\,{\rm M_\odot}$ in \hires, \fid and \norad respectively (Fig.~\ref{fig:icmf}). The fitted slopes are $\beta_{\rm tr}=-1.54\pm0.02$, $-1.34\pm0.05$ and $-1.60\pm0.05$. At the fiducial resolution, the ICMF is therefore shallower with stellar radiation. The Schechter form provides a viable description over the adopted fitting ranges. The pure-power-law fits give $p=0.015$, $0.0054$ and $0.013$ in \hires, \fid and \norad, respectively (Appendix~\ref{app:icmf}).

The slope of \hires is steeper than that of \fid by 0.19 and close to that of \norad, so convergence of the slope is not established. Above a common lower mass threshold of $10^4\,{\rm M_\odot}$, \fid remains shallower than \norad, although the contrast between these two runs depends on the exact lower fitting limit (Appendix~\ref{app:icmf}). The fitted high-mass cut-offs of the ICMF overlap, but their uncertainties preclude a precise test of feedback dependence. \hires forms 156 clusters against 100 in \fid above $10^5\,{\rm M_\odot}$, and 33 against 20 above $10^6\,{\rm M_\odot}$. We return to this abundance difference in Section~\ref{sec:clouds_mf}.

The main-galaxy ICMF is shallower than the slope of about $-2$ typical of young clusters in nearby normal star-forming galaxies \citep{portegieszwartYoungMassiveStar2010,krumholzStarClustersCosmic2019a,adamoStarClustersFar2020}, and close to the slopes of $-1.4$ to $-1.7$ found in local mergers and luminous infrared galaxies \citep{adamoStarClusterFormation2020,lindenMassiveStarCluster2021}, in all three runs. This contrasts with \citet{anderssonPresupernovaFeedbackSets2024a}, who find that in an isolated dwarf galaxy stellar radiation removes the most massive clusters and leaves a steep power-law ICMF. We discuss this difference in Section~\ref{sec:disc_sims}, and how stellar radiation acts between clouds and clusters in Section~\ref{sec:disc_icmf}. It is not clear, however, that the ICMF at cosmic dawn should match that of the local Universe, and high-redshift observed cluster populations are sparse. The pure-power-law slope of $-1.89$ measured for young, massive lensed clusters at $1<z<10$ \citep{claeyssensFirstGLIMPSEStar2026} is not directly comparable to the slope below the mass cut-off of our Schechter fits. Their sample is restricted to masses above $2\times10^6\,{\rm M_\odot}$, where curvature matters for our fitted $M_{\rm c}$. The local logarithmic slope of ${\rm d}N/{\rm d}M$ is $\beta_{\rm tr}-M/M_{\rm c}$, so the exponential factor already contributes about $-0.3$ at this lower value. The observed sample also pools galaxies over a wide redshift range, whereas our primary fit follows one galaxy. A fair quantitative comparison would require the same selection and fitted law. What is better measured at high redshift is the brightest cluster belonging to a galaxy, which samples the upper end of the ICMF, and in the next section we use this statistic to test our simulations.

%==============================================================================
\subsection{The brightest clusters and the host star formation rate}
\label{sec:brightest}

\begin{figure*}
    \includegraphics[width=\textwidth]{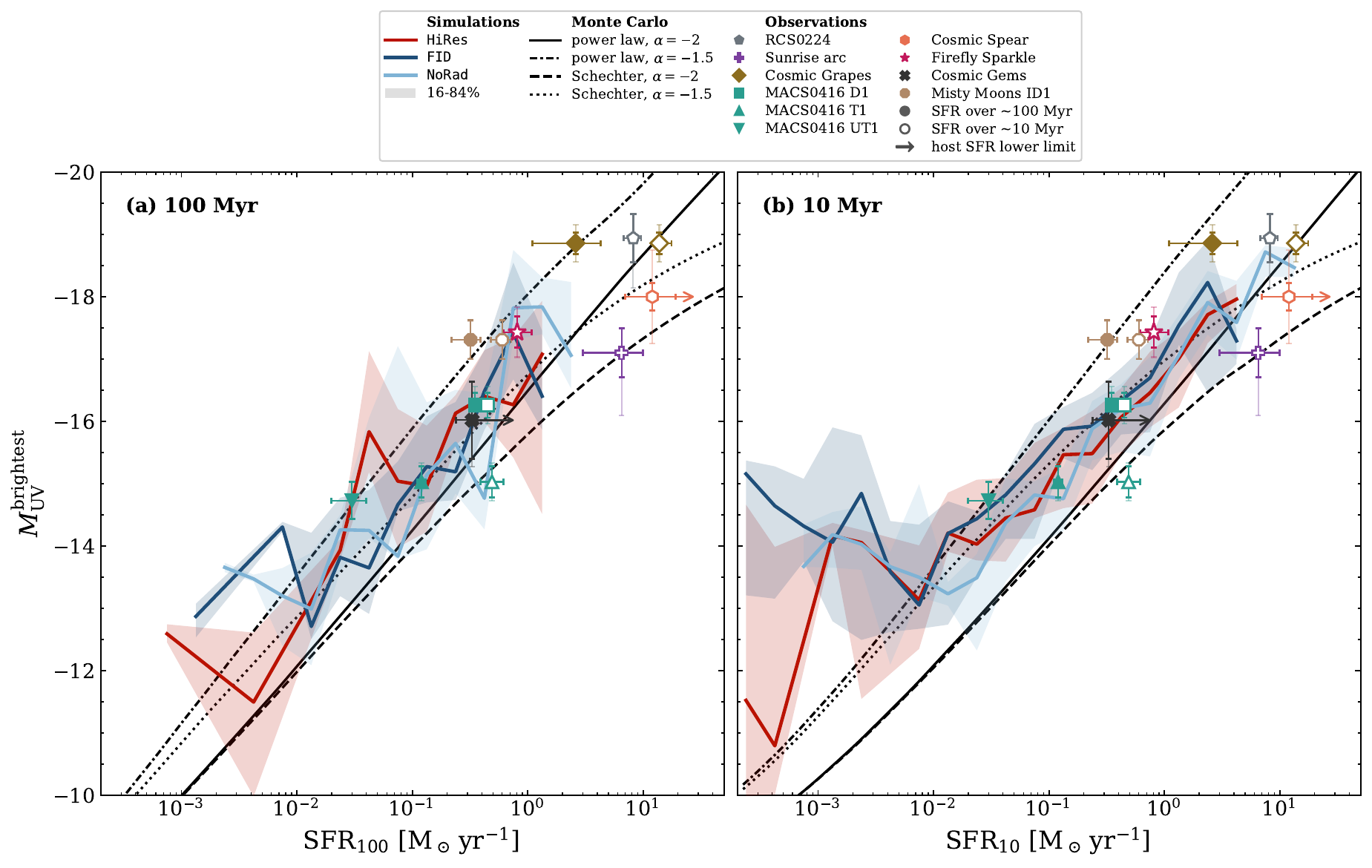}
    \vspace{-5mm}
    \caption{Rest-frame 1500\,\AA\ absolute magnitude of the brightest cluster of the main galaxy against the galaxy's SFR, averaged over the preceding 100~Myr (left) and 10~Myr (right). For \hires, \fid and \norad (identified in the legend), thick lines give the median magnitude in 0.25\,dex bins of SFR, drawn for bins holding at least 5~Myr of star formation history, and shaded bands the 16th--84th percentiles of those instants (see text for details). Symbols show the lensed systems of Table~\ref{tab:obs}, one per system and per SFR tracer, on both panels, filled for the published SFR averaged over about 100~Myr and open for the SFR averaged over about 10~Myr, with arrows marking lower  limits. Darker error bars are the random errors, from photometry and the statistical uncertainty of the magnification, and fainter error bars indicate lens-model systematics, one-sided towards fainter magnitudes for the Sunrise arc, whose magnifications are lower limits, and for the Cosmic Gems arc, for which the adopted lens model gives the lowest of three magnifications, asymmetric for RCS0224 as in the source paper, and symmetric otherwise. Thin lines show the median brightest cluster expected from Monte Carlo sampling of pure power-law ICMFs with $\alpha=-2$ (solid) and $-1.5$ (dash-dotted), and of Schechter ICMFs with $M_{\rm c}=7\times10^6\,{\rm M_\odot}$ and $\alpha=-2$ (dashed) or $-1.5$ (dotted), assuming all stars form in clusters. Clusters are selected within $0.2R_{\rm vir}$ at observation and include all attached stellar populations. Magnitudes and Monte Carlo curves include stellar light only, without nebular emission or dust.}
    \alttext{Two panels plot the ultraviolet magnitude of the brightest cluster against host star formation rate, averaged over one hundred million years on the left and ten million years on the right. Higher star formation rates generally accompany brighter clusters. Simulated tracks and observed lensed systems are compared with four mass-function sampling predictions. At low short-timescale rates, the simulated tracks retain bright clusters and curve away from the predictions.}
    \label{fig:brightest}
    \vspace{-3mm}
\end{figure*}

The brightest cluster in a galaxy of given SFR is one of the few statistics of the cluster population that can be measured at high redshift \citep{bastianStarFormationRate2008,adamoProbingRoleGalactic2015,pfefferComparingEMOSAICSPredictions2025}. Figure~\ref{fig:brightest} compares our runs with the observed lensed galaxies listed in Table~\ref{tab:obs}. Every 0.5~Myr we select the brightest cluster currently inside the central aperture, whatever its age or birth location. Its light includes all attached stellar populations, including those gained through mergers or renewed star formation. We pair it with the SFR of the host galaxy averaged over the preceding 100 or 10~Myr. The tracks give median magnitudes in 0.25-dex SFR bins containing at least ten 0.5~Myr instants (i.e. 5 Myr of star formation history across the entire redshift range spanned by the simulations), and therefore weight each SFR by the amount of time the galaxy spends forming stars at this rate. Only 57, 39 and 40 distinct clusters are ever brightest in \hires, \fid and \norad. Observed SFRs are shown on both panels, as filled symbols for SFR tracers sensitive to $\sim$100~Myr and open symbols for those probing $\sim$10~Myr. Quoted offsets use only the matching tracers. Arrows mark host SFR lower limits. Darker magnitude error bars show photometric and magnification uncertainties, while fainter bars show lens-model systematics, which we do not propagate into the offsets (Appendix~\ref{app:obs}).

To interpret these brightest cluster data, we compute the luminosity of the brightest cluster that we would expect from randomly sampling a given analytic ICMF \citep{bastianStarFormationRate2008,adamoProbingRoleGalactic2015}, which we refer to as the Monte Carlo prediction. In other words, for each SFR bin, we draw clusters between $10^3$ and $10^9\,{\rm M_\odot}$ until their total mass equals the stellar mass formed in the SFR time window, assuming that all stars form in clusters. We assign them ages uniformly over the time window and metallicities drawn from the simulated cluster population, and record the brightest. We use four reference ICMFs, ${\rm d}N/{\rm d}M\propto M^{\alpha}$ with $\alpha=-2$ or $-1.5$, either pure or truncated at $M_{\rm c}=7\times10^6\,{\rm M_\odot}$, compatible with the fitted Schechter mass cut-offs (Section~\ref{sec:icmf}). We measure the offset of a sample from each Monte Carlo prediction as the median -- either over all the SFRs in a bin for the simulations or over all the systems in that SFR bin for the observations -- of the difference between the magnitude of the brightest cluster and its predicted magnitude, and report it in terms of magnitudes above or below the prediction.

Using the nominal $\sim$100~Myr averaged SFR estimates, both the observed and simulated samples lie above the pure $\alpha=-2$ power law ICMF Monte Carlo reference. The brightest clusters of the three runs lie 0.6--0.7\,mag above its median prediction, compared with about 1.1\,mag for the six lensed systems with a published SFR$_{100}$. These offsets exclude lens-model systematics and dust-correction uncertainties (see Appendix~\ref{app:obs}). The Monte Carlo curves are medians, however, and the brightest cluster of a randomly sampled population is drawn from the sparsely populated upper mass end of the ICMF, so its magnitude varies widely between realizations. For $\alpha=-2$, half of the observed points on the 100~Myr panel and most of those on the 10~Myr panel lie within the 16th--84th percentile range of the brightest cluster, and all of them within the 2.5th--97.5th percentile range. Within that spread the observed points are consistent with an $\alpha=-2$ power law but also, on the 100~Myr panel, with a shallower, truncated ICMF. Such an ICMF predicts a narrower spread, and the same comparison applied to our simulations, whose ICMFs we know, does not recover them consistently. On its own, the brightest cluster at a given SFR therefore cannot clearly distinguish between ICMFs, and the three runs have similar median offsets at this time-scale.

We also note that the simulated tracks end at $\mathrm{SFR}_{100}$ values of 1 to $3\,{\rm M_\odot\,yr^{-1}}$, which are smaller than those of half of the hosts with a measured short-time-scale SFR (left panel). However, when
averaged over 10~Myr, the simulated SFRs peak at several times their 100~Myr value (see Fig.~\ref{fig:sfr}), and the tracks now reach the short-time-scale observations (right panel). The observed hosts move the same way, since every system with an SFR measured on both time-scales has a higher SFR when averaged over 10~Myr than over 100~Myr. This is a signature of bursty star formation, and a galaxy selected to host a bright young cluster is likely to be caught during a burst. On the 10~Myr panel, the simulated brightest clusters of any age lie further above the $\alpha=-2$ median prediction than when the SFR is averaged over 100~Myr, and the tracks lift off the relation at low $\mathrm{SFR}_{10}$, whereas the lensed clusters with a measured $\sim$10~Myr SFR scatter about the median curve. Most of the observed clusters are younger than about 15~Myr (see Table~\ref{tab:obs}), so they likely formed during the episode of star formation that the short-time-scale SFR of their host measures, whereas over 100~Myr the same burst is diluted. As a result, observed cluster data points move horizontally towards lower SFRs and end up above the median prediction. As \citet{bastianStarFormationRate2008} sketched for starbursts, a galaxy and its brightest cluster move together along the relation while the galaxy forms clusters. Once the burst ends, the SFR averaged over 10~Myr drops significantly within a few Myr, but the brightest cluster fades more slowly, so the galaxy moves to lower SFR at almost constant cluster brightness, above the predicted relation. Our bursty galaxies spend much of their time in this post-burst state, which is short compared with a 100~Myr averaging window, so the offset is smaller on that longer time-scale. Any comparison with observations therefore depends on how fast bright clusters fade and on the time-scale to which the chosen SFR tracer responds (see Section~\ref{sec:disc_obs}).

\section{Discussion}
\label{sec:discussion}

We now examine how stellar radiation changes the gas supply and duration of cluster growth, and how these shape the ICMF.

%==============================================================================
\subsection{What sets the initial cluster mass function}
\label{sec:disc_icmf}

The ICMF depends on the population of available clouds as well as on the clusters they produce (see Section~\ref{sec:intro}). Our runs provide a measurement of both, either with or without radiative feedback. A cloud can form several clusters, so its contribution to the ICMF depends on both their number and their growth. All Pop~II stars form through sink particles, but the fraction that remains bound is not modelled (Section~\ref{sec:limits}). Our cloud census and natal-cloud sample are also built differently (Section~\ref{sec:clouds_where}).

\subsubsection{The natal clouds}
\label{sec:clouds}
\label{subsec:clouds}
\label{sec:clouds_mf}

We identify structures in post-processing using \textsc{phew} \citep{bleulerPHEWParallelSegmentation2015} above the hydrogen density thresholds used by \citet{quinalhaStellarFeedbackEffects2026}, $n_{\rm H}>10$, $10^{1.5}$ and $100\,{\rm cm^{-3}}$, which are referred to as cloud complexes, big clouds and clouds. We merge all substructure, so each structure is a connected region above its threshold. In the remainder of this paper, `clouds' means structures with $n_{\rm H}>100\,{\rm cm^{-3}}$ unless otherwise stated, and we analyse structures above each density threshold separately.

\begin{figure*}
    \centering
    \includegraphics[width=\textwidth]{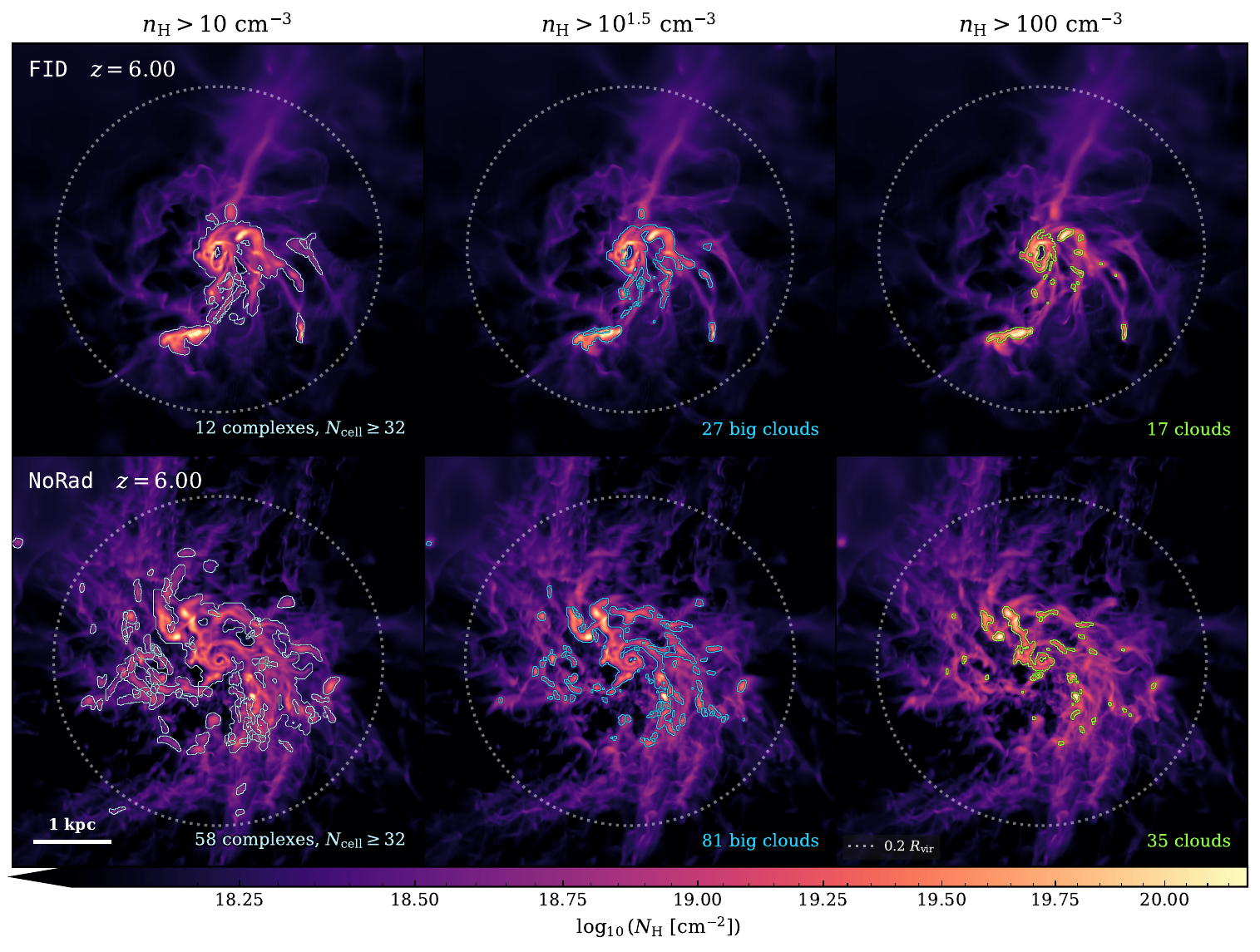}
    \vspace{-5mm}
    \caption{Hydrogen column density of the main galaxy at $z=6$ in \fid (top row) and \norad (bottom row). Each panel is $5.25\,{\rm kpc}$ across. Contours outline connected structures above $n_{\rm H}>10$ (complexes, left), $10^{1.5}$ (big clouds, centre) and $100\,{\rm cm^{-3}}$ (clouds, right). Complexes are drawn if they contain at least 32 cells, and denser structures are nested inside them. The number of structures drawn appears in the lower right of each panel. Dotted circles delineate regions of radius $0.2\,R_{\rm vir}$, and the bar in the bottom left corner of the figure gives the physical scale. The column-density scale is shared by both rows.}
    \alttext{Six hydrogen column-density maps compare FID above with NoRad below at redshift six. Columns outline successively denser complexes, big clouds and clouds. Smaller dense structures are nested inside larger ones. NoRad contains more outlined structures at every threshold. Dotted circles mark the central galaxy aperture, and brighter colours indicate higher column density.}
    \label{fig:ism}
    \vspace{-3mm}
\end{figure*}

\begin{figure*}
    \centering
    \includegraphics[width=\textwidth]{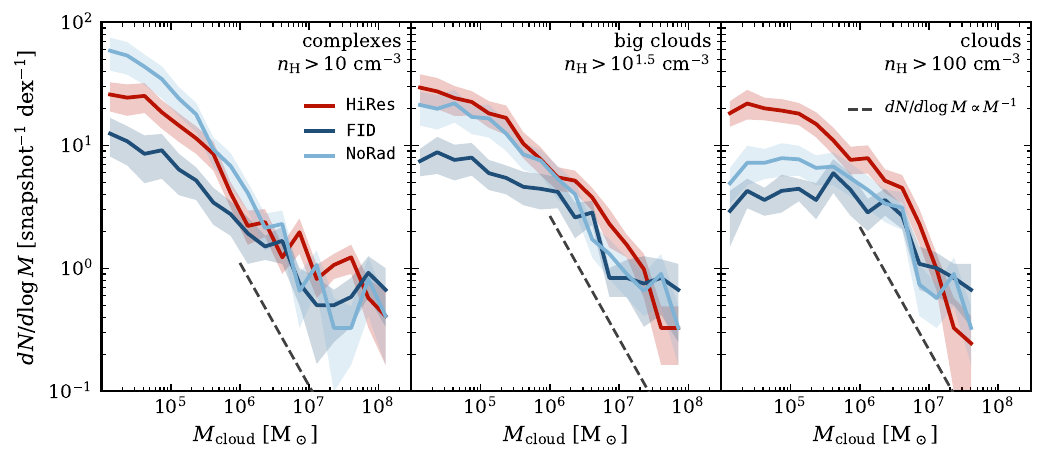}
    \caption{Mass functions, as detections per snapshot per dex, of structures whose density peaks lie within the central $0.2R_{\rm vir}$ aperture, for snapshots spaced every 10 Myr between $z=12$ and 6, for complexes ($n_{\rm H}>10\,{\rm cm^{-3}}$, left), big clouds ($10^{1.5}\,{\rm cm^{-3}}$, centre) and clouds ($100\,{\rm cm^{-3}}$, right). The census pools snapshots spaced $10\,{\rm Myr}$ apart in \hires, \fid and \norad (identified in the legend), and a persistent structure contributes at each detection. Lines show the pooled counts and bands the 16th--84th percentiles of a block bootstrap over snapshots within five-snapshot ($50\,{\rm Myr}$) blocks. All structures are included, whatever their number of cells, and bins with fewer than three detections are omitted. A structure of 32 cells at each of the three threshold densities has a mass of $4.5\times10^4$, $1.4\times10^5$ and $4.5\times10^5\,{\rm M_\odot}$ in \fid and \norad (at the $z=6$ cell size) and one eighth of these values in \hires, so counts below these masses are incomplete. Note that the cell size, and with it these masses, varies over the redshift range of the census (see Section~\ref{subsec:initial_conditions}). The dashed lines indicate ${\rm d}N/{\rm d}\log M\propto M^{-1}$, equivalent to ${\rm d}N/{\rm d}M\propto M^{-2}$, and have arbitrary normalizations.}
    \alttext{Three logarithmic panels show mass functions of complexes, big clouds and clouds, from left to right. Counts generally decrease towards larger masses. NoRad has more low-mass complexes, while HiRes has more dense clouds than FID over much of the plotted range. Shaded bands show uncertainty, and dashed reference lines indicate a minus-one logarithmic slope.}
    \label{fig:comf}
\end{figure*}

Figure~\ref{fig:ism} illustrates the gas structures detected at $z=6$ in the \fid and \norad simulations. Both runs contain a large central complex, while the surrounding gas in \norad is broken into more structures. The mass functions plotted in Fig.~\ref{fig:comf} sample these structures over the assembly history of the main halo. More specifically, we pool simulation snapshots taken every 10\,Myr from $z=12$ to 6, selecting structures whose density peaks lie within $0.2R_{\rm vir}$. Persistent structures contribute once per snapshot, as in the stacked census of \citet{quinalhaStellarFeedbackEffects2026}. We plot the number of detections per snapshot per dex, including all structures, regardless of how many cells they contain. A block bootstrap over consecutive snapshots accounts for some of the temporal correlation between detections. These mass functions thus describe a time-sampled population, rather than a census of distinct structures.

Figure~\ref{fig:comf} shows two regimes. Below $10^6\,{\rm M_\odot}$ the three runs differ at every density threshold. \norad holds several times as many complexes per snapshot as \fid, and its excess shrinks with increasing density but does not vanish for the dense clouds. \hires holds more structures than \fid at every threshold, most of all among the low-mass dense clouds, which the coarser grids resolve with only a few tens of cells. Above $10^6\,{\rm M_\odot}$ the mass functions approach one another, with overlapping bootstrap bands in some bins. There the numbers are small, a few detections per snapshot in the most massive bins and often of the same persistent structure, so we cannot distinguish a feedback effect at the upper end of the mass functions. No single power law describes these upper ends, but slopes fitted above $10^6\,{\rm M_\odot}$ are shallower with stellar radiation than without it for the complexes of \hires and \fid and for the dense clouds of \fid.

The low-mass counts must be interpreted relative to the resolution scales in Fig.~\ref{fig:comf}. At $z=6$, a 32-cell structure at the threshold density has a mass of $4.5\times10^4$, $1.4\times10^5$ and $4.5\times10^5\,{\rm M_\odot}$ for complexes, big clouds and dense clouds respectively, in both \fid and \norad. The corresponding masses in \hires are one eighth of these values. These scales vary with the physical cell size during the census and are indicative rather than sharp completeness limits. The matched-resolution \fid--\norad comparison isolates the stellar-radiation switch, while comparisons with \hires also change how well structures are resolved.

Stellar radiation changes the cloud population most strongly below $10^6\,{\rm M_\odot}$ and at low density. The finer grid also increases low-mass counts by following fragmentation further before sink particle creation. This contributes to the larger cluster abundance in \hires, where gas accreted by one sink particle in \fid can instead be shared among several. The radiation-on/off comparison at fixed resolution therefore provides the clearest test to isolate the effect of feedback.

In an isolated dwarf, \citet{quinalhaStellarFeedbackEffects2026} found that complexes become more top-heavy with stellar radiation while the dense cloud mass function keeps its shape. As in \citet{quinalhaStellarFeedbackEffects2026}, we do observe that the feedback contrast weakens at higher density thresholds in our simulations. However, our dense-cloud counts still differ at low mass, while the small number of massive structures prevents a strong test of feedback independence at the upper mass end. These authors' mass functions are also cut off below $\sim10^{6.4}\,{\rm M_\odot}$, whereas ours extend to $10^8\,{\rm M_\odot}$, and the different environments and star formation prescriptions prevent us from attributing the contrast to a unique, well-defined cause.

\subsubsection{Where clusters form}
\label{sec:clouds_where}

We link each sink particle to its birth cloud by tracking its birth position back to the time of the preceding snapshot (typically $2\,{\rm Myr}$, and at worst $4\,{\rm Myr}$) using a velocity fitted to its early trajectory. If the sink particle's back-projected birth position lies within an isodensity contour defining a cloud, it is assigned that cloud. Failing that, we assign the cloud that contains the sink particle at the first output after its formation. Clusters born in the same cloud are \emph{co-natal}, and a cloud that forms at least one cluster is \emph{natal}. We call a cluster \emph{unlinked} if neither match succeeds. This can occur when clusters form in gas below the density threshold considered, or when clouds evolve rapidly between successive outputs. Matches use every output and are retained even if the natal cloud lies beyond the $0.2R_{\rm vir}$ aperture.

Radiative feedback also changes where sink particles form. At $n_{\rm H}>100\,{\rm cm^{-3}}$, 99, 79 and 43~per~cent of the cluster birth sample in \hires, \fid and \norad are respectively linked to a cloud. Without stellar radiation, more sink particles are therefore not linked to structures selected by this density threshold. The lower natal-cloud matching fraction in \norad is consistent with sink formation in less dense gas. In \norad, the colder seeding gas lowers the density required by the sink particle formation criterion, which scales as $c_{\rm s}^2/\Delta x_{\rm min}^2$ (see equation~\ref{eq: LP condition}). The temperature dependence of the seeding criterion provides a second route by which stellar radiation changes cluster formation, alongside its effect on the cloud population.

\subsubsection{Converting clouds into clusters}
\label{sec:conversion}

\begin{figure}
    \includegraphics[width=\linewidth]{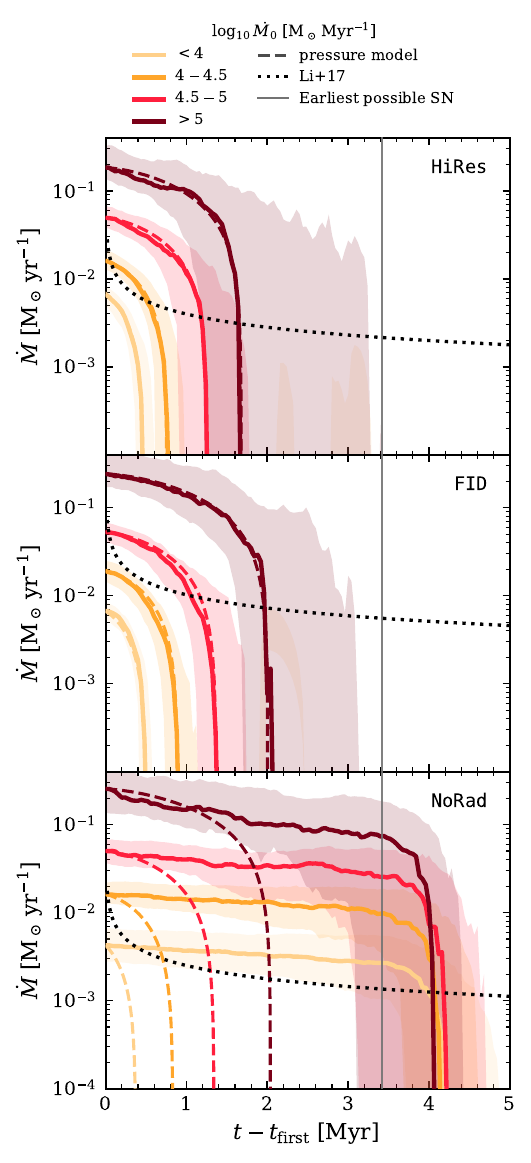}
    \vspace{-5mm}
    \caption{Median sink particle accretion rate against time since the first star particle formed, $t-t_{\rm first}$, in bins of initial accretion rate, $\log_{10}[\dot M_0/({\rm M_\odot\,Myr^{-1}})]<4$, 4--4.5, 4.5--5 and $>5$ (legend at the top). Rows show \hires, \fid and \norad, with shaded 25th--75th percentile ranges. Zero accretion rates are drawn below the axis. Dashed lines show equation~(\ref{eq:Mdot}) with $p=1/2$ fitted to each \hires and \fid median track, ending at accretion shut-off. The \norad panel uses \fid's fitted coupling in each initial-rate bin with \norad's own initial rate for reference and is not fitted. The dotted line indicates the approximate $\dot M\propto t^{-0.5}$ decline found by \citet{liStarClusterFormation2017} in their 10~per~cent efficiency-per-free-fall-time run, normalized to each panel's median early rate. The vertical line marks the earliest possible supernova, at the minimum model delay of $\sim3.4\,{\rm Myr}$ after $t_{\rm first}$, the same in all three runs.}
    \alttext{Three stacked panels show median cluster accretion histories, grouped by initial accretion rate, for HiRes, FID and NoRad. In the radiation runs, accretion drops sharply to zero, with higher-rate groups generally lasting longer. NoRad sustains accretion until around four million years. Dashed model tracks and a vertical earliest-supernova marker provide comparisons.}
    \label{fig:acctracks}
    \vspace{-3mm}
\end{figure}

Figure~\ref{fig:acctracks} shows the median accretion rate of sink particles, binned by their initial accretion rate $\dot M_0$ at $t_{\rm first}$, against the time elapsed since they created their first sub-grid stellar particle, $t-t_{\rm first}$. With stellar radiation turned on, the accretion rate declines from this point and reaches zero before the first supernovae associated with the sink particle go off. Sink particles with higher initial accretion rates generally accrete for longer. In \norad the accretion rate stays closer to a plateau, across the initial-rate bins, until the first supernovae explode, and then falls abruptly.

The declining accretion rates motivate an empirical braking law for \hires and \fid, in which the rate of decline increases as the stellar population grows.
We first write this law with a general exponent $p$, as:
\begin{equation}
    \frac{{\rm d}^2 M_\star}{{\rm d}t'^2}=-K M_\star^{p},
    \label{eq:pressure_model}
\end{equation}
where $t'=t-t_{\rm first}$, $M_\star(0)=m_{\rm p}=10^3\,{\rm M_\odot}$ is the first star particle created, and $\dot M_\star(0)=\dot M_0$. The coupling constant, $K$, absorbs the effects of geometry, gravity and gas supply. We stop the solution at the cluster formation time, $t_{\rm form} \equiv t_{\rm end}-t_{\rm first}$, the first zero of $\dot M_\star(t')$, where $t_{\rm end}$, should the model accurately describe the simulations, marks the end of the first gas accretion episode on to the sink particle. Here $M_\star(t')$ is a smooth approximation to the stellar mass build up in the simulation, which is recorded in the form of discrete stellar particles. Integrating once with respect to time gives:
\begin{equation}
    \dot M_\star^2=\dot M_0^2-\frac{2K}{p+1}
    \left(M_\star^{p+1}-m_{\rm p}^{p+1}\right),
    \label{eq:Mdot}
\end{equation}
and the final mass is:
\begin{equation}
    M_{\star,\rm end}=\left(m_{\rm p}^{p+1}+\frac{(p+1)\dot M_0^2}{2K}\right)^{1/(p+1)}.
    \label{eq:mstar_pressure}
\end{equation}
For $p>-1$, the coupling has dimensions $[K]={\rm mass}^{1-p}\,{\rm time}^{-2}$. At fixed $p$ and with $K$ independent of $\dot M_0$, the limit $M_{\star,\rm end}\gg m_{\rm p}$ gives $M_{\star,\rm end}\propto\dot M_0^{2/(1+p)}$ and
$t_{\rm form}\propto\dot M_0^{(1-p)/(1+p)}$. The duration scaling follows by integrating ${\rm d}t'={\rm d}M_\star/\dot M_\star$ from $t'=0$ to $t'=t_{\rm form}$. Both exponents therefore depend on the adopted braking law. We further retain $p=1/2$ as a simple empirical description of the accretion histories (see Appendix~\ref{app:feedback_models} for a brief discussion of the reasons for this choice), giving the asymptotic scalings $M_{\star,\rm end}\propto\dot M_0^{4/3}$ and $t_{\rm form}\propto\dot M_0^{1/3}$. The law describes the broad decline of the median accretion tracks in \hires and \fid (see Fig.~\ref{fig:acctracks}), but does not identify which radiative process brakes the inflow. The large-mass approximation fails near the single-particle scale, and the measured duration relation is curved (see Section~\ref{sec:chain}). We do not fit \norad, whose sink particles generally accrete until their first supernovae explode, and use \fid's fitted coupling in each initial-rate bin with \norad's own initial rate for the reference curves in the bottom panel of Fig.~\ref{fig:acctracks}.

Different sink particles within the same bin stop accreting at different times, so a median track is not the history of a single sink particle. We therefore calibrate $K$ on individual tracks, with $\dot M_0$ fixed to the measured rate at time $t_{\rm first}$ (see Appendix~\ref{app:feedback_models} for details). Individual-track fits give median couplings of 281 and $260\,{\rm M_\odot^{1/2}\,Myr^{-2}}$ in \hires and \fid, with 16th--84th percentile ranges of 172--416 and $154$--$371\,{\rm M_\odot^{1/2}\,Myr^{-2}}$ respectively. Similar couplings describe the simulations at the two resolutions. The most massive sink particles scatter more about the model, because fresh gas sometimes reaches them late during their growth (see Appendix~\ref{app:chain}). We therefore regard this as an empirical growth model.

\begin{figure}
    \includegraphics[width=\linewidth]{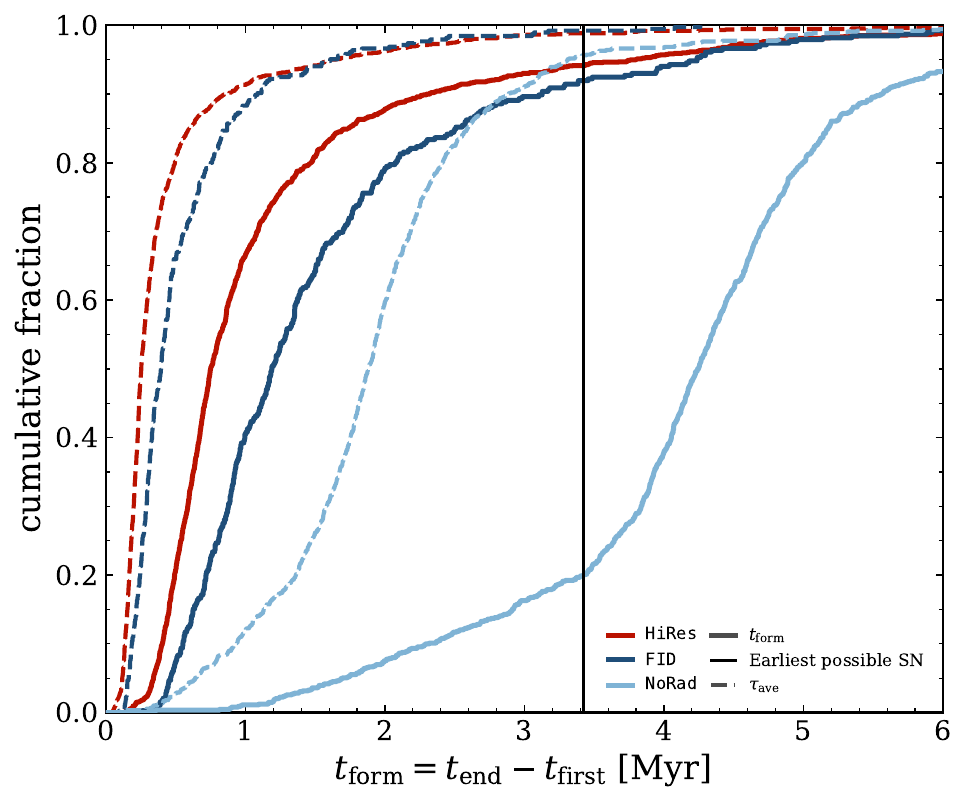}
    \vspace{-5mm}
    \caption{Cumulative distributions of the formation time $t_{\rm form}=t_{\rm end}-t_{\rm first}$ (solid), from the first star particle to the final permanent zero of the accretion rate, and of the accretion-weighted formation time $\tau_{\rm ave}$ (dashed), in \hires, \fid and \norad, for the birth sample, without a birth-mass cut. The vertical line marks the earliest possible supernova, at the minimum model delay of $\sim3.4\,{\rm Myr}$ after $t_{\rm first}$, the same in all three runs.}
    \alttext{Cumulative distributions compare total and accretion-weighted cluster formation times. HiRes has the shortest total durations, FID is intermediate, and NoRad has the longest. Most radiation-run episodes finish before the earliest possible supernova, whereas most NoRad episodes finish later. Dashed accretion-weighted distributions lie to the left of the corresponding solid total-duration curves.}
    \label{fig:tform}
    \vspace{-3mm}
\end{figure}

In addition to the cluster formation time, $t_{\rm form}$, we calculate the accretion-weighted mean time, $\tau_{\rm ave}=\int_{t_{\rm first}}^{t_{\rm end}}(t-t_{\rm first})\,\dot M\,{\rm d}t/\!\int_{t_{\rm first}}^{t_{\rm end}}\dot M\,{\rm d}t$, from the same time origin. Figure~\ref{fig:tform} shows both time-scales for the cluster birth sample. Cluster median formation times are 0.75, 1.19 and $4.24\,{\rm Myr}$, with median accretion-weighted times of 0.25, 0.39 and $1.88\,{\rm Myr}$ in \hires, \fid and \norad respectively. \citet{liStarClusterFormation2018} observed a similar trend, with mass-weighted formation times shorter than their unweighted counterparts. Most of the growth thus occurs early when simulations account for stellar radiation, and slightly earlier still at the finer resolution. In \hires and \fid, 94.2 and 92.0~per~cent of the main-galaxy sink particles, respectively, stop accreting before the minimum supernova delay of $\sim3.4\,{\rm Myr}$ after $t_{\rm first}$, compared to only 20.0~per~cent in \norad. This places the termination of most radiating cluster accretion episodes before their own stellar populations produce a supernova, although supernovae from neighbouring sink particles can already have exploded. With stellar radiation, sink particles also accrete for longer when located in more massive natal clouds. Indeed, the Spearman rank correlation coefficients between $t_{\rm form}$ and natal-cloud mass are $\rho=0.55$, 0.65 and $-0.07$ in \hires, \fid and \norad respectively. More massive clouds end up forming more massive star clusters in every run, but the relation between cloud mass and growth duration changes with stellar radiation, which is critical in mapping the natal-cloud population to the ICMF (Section~\ref{sec:chain}).

\begin{figure}
    \includegraphics[width=\linewidth]{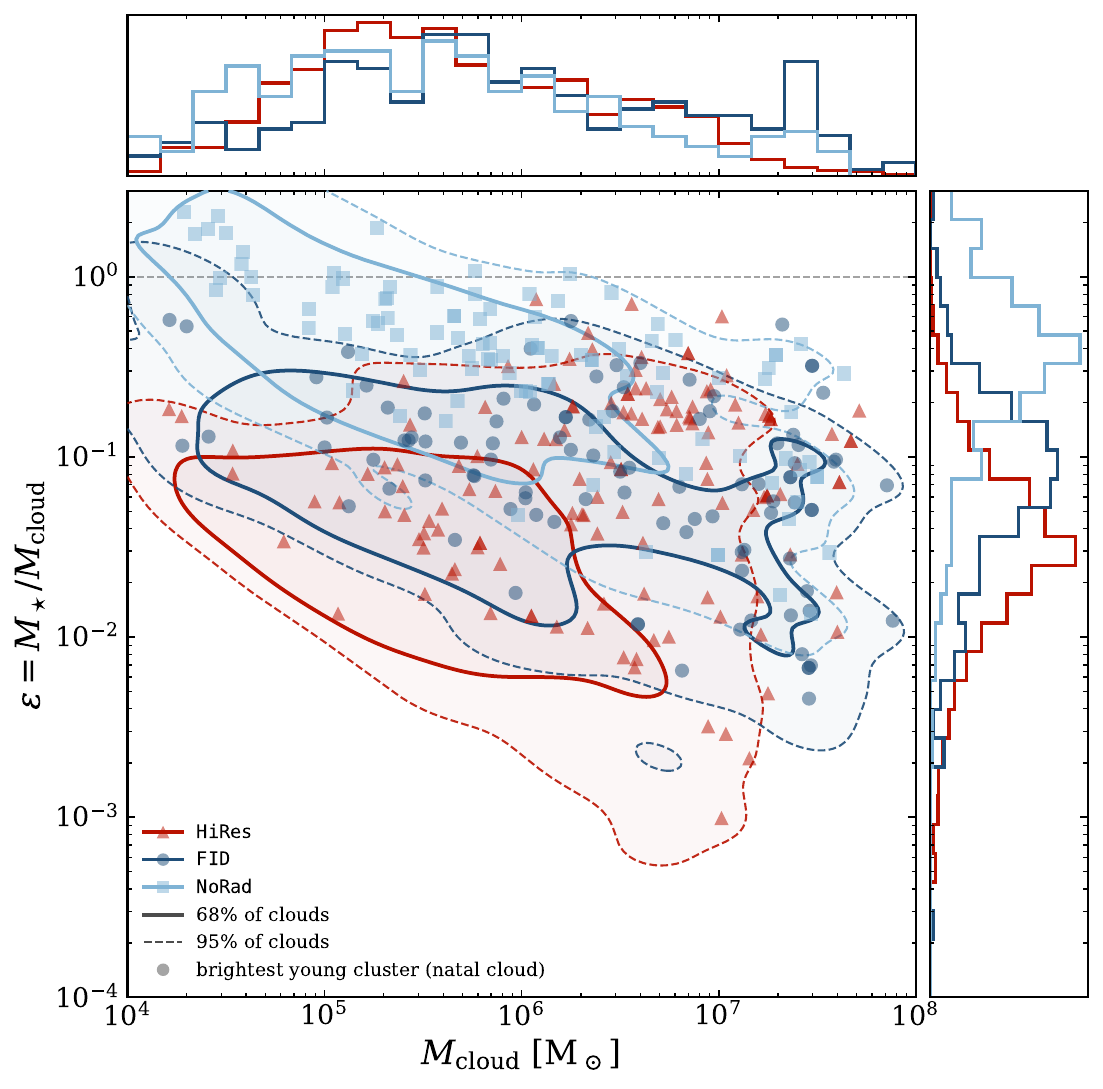}
    \vspace{-5mm}
    \caption{Integrated stellar-to-cloud mass ratio $\epsilon=\sum M_{\star,\rm birth}/M_{\rm cloud}$ against matched natal-cloud mass for $n_{\rm H}>100\,{\rm cm^{-3}}$ in \hires, \fid and \norad. The samples contain 1090, 246 and 331 natal clouds in these 3 runs respectively, and birth masses of co-natal sink particles in the birth sample are summed. The cloud mass is measured at the matched snapshot and $\epsilon$ is not capped at unity (horizontal dashed line, see text for details). Contours enclose 68 (solid) and 95~per~cent (dashed) of each sample, with marginal distributions shown above and to the right of the central panel. Triangles, circles and squares mark the natal clouds of the clusters of \hires, \fid and \norad, respectively, that are at some instant (sampled every 0.5~Myr) the brightest of those formed within the preceding 10 or 100~Myr, using the same observation-time aperture as Fig.~\ref{fig:brightest}. A symbol is shown only when the matched cloud is also included in the natal-cloud sample used for the contours.}
    \alttext{Contour distributions plot the integrated stellar-to-cloud mass ratio against natal-cloud mass, with marginal histograms above and to the right. NoRad is generally shifted to higher ratios than FID, and HiRes to lower ratios. Some ratios exceed the horizontal unity line. Symbols identify natal clouds associated with the brightest young clusters.}
    \label{fig:eps}
    \vspace{-3mm}
\end{figure}

Radiative feedback also lowers the stellar mass formed relative to the matched natal-cloud mass (Fig.~\ref{fig:eps}). We define
\begin{equation}
    \epsilon=\frac{\sum M_{\star,\rm birth}}{M_{\rm cloud}},
    \label{eq:eps}
\end{equation}
summing over co-natal sink particles in the main-galaxy birth sample. In \hires, \fid and \norad\ the median $\epsilon$ is 0.031, 0.078 and 0.36, with 16th--84th percentile ranges of 0.010--0.082, 0.020--0.19 and 0.10--1.05 respectively. We do not cap this ratio at unity. Values above unity, found for 56 of the 331 natal clouds of \norad, one of 1090 in \hires and five of 246 in \fid, mean that these sink particles ended up forming more stellar mass than their cloud held in gas when it was matched, consistent with continued gas supply from outside that initial reservoir. $\epsilon$ is therefore not the conversion efficiency of a closed system. Note that $\epsilon$ is only calculated for the cloud-linked sample, which includes a smaller fraction of clusters in \norad. In every run, $\epsilon$ rises with cloud mass but, as more massive clouds host more sink particles, this does not mean that a single sink particle is more efficient at forming a cluster because it is located in a more massive cloud (see Appendix~\ref{app:chain}). The factor of a few between \fid and \norad at fixed cloud mass is the integrated effect of the shorter accretion episodes of Fig.~\ref{fig:tform}, and the factor of two to three between \fid and \hires follows from the lower initial accretion rates of \hires at fixed cloud mass (see Appendix~\ref{app:chain}), with the same coupling. The natal clouds marked for hosting the brightest young clusters in Fig.~\ref{fig:eps} span the measured range of cloud masses and, with stellar radiation, lie mostly above the median $\epsilon$ of all natal clouds.

%==============================================================================
\subsection{Stellar radiation and the end of accretion}
\label{sec:shutdown}
\label{subsec:shutdown_method}
\label{sec:disc_resolution}

The comparison with \norad shows that stellar radiation significantly shortens these accretion episodes. We compare the expected roles of photoionization heating and direct radiation pressure using pressure-balance radii \citep{rosdahlSchemeRadiationPressure2015,kimmFeedbackregulatedStarFormation2017,hanImpactRadiationFeedback2022a}, given by:
\begin{equation}
r_{\rm PH}=\left(\frac{3Q}{4\pi\alpha_{\rm B}}\right)^{1/3}
\left(\frac{k_{\rm B}T_{\rm ion}}{P_{\rm conf}}\right)^{2/3}
\label{eq:radius_photoheating}
\end{equation}
for photoionization heating and
\begin{equation}
r_{\rm DP}=\left(\frac{L_{\rm abs}}{4\pi cP_{\rm conf}}\right)^{1/2}
\label{eq:radius_direct_pressure}
\end{equation}
for direct radiation pressure. In these equations, $Q$ is the stellar ionizing photon rate and $L_{\rm abs}$ the absorbed luminosity. We adopt an ion temperature $T_{\rm ion}=10^4\,{\rm K}$ and the case-B recombination coefficient $\alpha_{\rm B}=2.59\times10^{-13}\,{\rm cm^3\,s^{-1}}$, with $k_{\rm B}$ the Boltzmann constant and $c$ the speed of light. At each available snapshot between $t_{\rm first}$ and $t_{\rm end}$, we calculate individual sink particle radiative outputs by summing over the stellar populations they have already formed, taking into account masses, ages and metallicities, and interpolating from \textsc{bpass} SEDs. We set $L_{\rm abs}$ to the six-group luminosity emitted in the simulations as a full-absorption limit.

The definition of the confinement pressure, $P_{\rm conf}$, requires care at our resolution. \citet{kimmFeedbackregulatedStarFormation2017} estimate ambient thermal pressure from neighbouring cells, while \citet{hanImpactRadiationFeedback2022a} include thermal, turbulent and gravitational terms across a resolved cloud core. Our accretion region spans only one finest-cell radius, so it does not generally allow us to separate the ambient cloud from the gas photo-heated by the sink particle. Including that heated gas pressure as a source of confinement would be counting the same thermal feedback as both the driving pressure and the resistance to expansion. Neglecting ambient thermal pressure (see fig.~6 of \cite{hanImpactRadiationFeedback2022a} and discussion at the end of this section for a justification), we therefore use $P_{\rm conf}=P_{\rm grav}+P_{\rm ram}$, with:
\begin{equation}
\begin{aligned}
P_{\rm grav}&=-\langle\rho\boldsymbol{g}\cdot\boldsymbol{r}\rangle_V,\\
P_{\rm ram}&=\langle\rho[\min(v_r,0)]^2\rangle_V.
\end{aligned}
\label{eq:confinement}
\end{equation}
We sample the sink particle-centred sphere of radius $r\leq r_{\rm acc}=\Delta x_{\rm min}$ on a uniform lattice of spacing $\Delta x_{\rm min}/4$, assigning to each point the density, $\rho$, velocity, $\boldsymbol{v}$, and total gravitational acceleration, $\boldsymbol{g}$, of its enclosing AMR leaf cell. Both terms are averaged over the full selected volume. Here $\boldsymbol{r}$ denotes position relative to the sink particle and the relative radial velocity is thus $v_r=(\boldsymbol{v}-\boldsymbol{v}_{\rm sink})\cdot\hat{\boldsymbol{r}}$. We exclude outflowing gas when calculating  $P_\mathrm{ram}$, while the average remains over the full sampled volume. This volume-averaged inward kinetic stress provides an effective resistance to expansion.

A pressure-balance radius exceeding $r_{\rm acc}$ indicates that the channel can pressurise the whole accretion region under the assumed confinement. For a given $Q$, a larger ionized volume has a lower equilibrium density and therefore lower thermal pressure. The radius $r_{\rm PH}$ estimates how far photoheated gas could expand before its thermal pressure balances $P_{\rm conf}$, rather than how much of the dense inflow is actually ionized. Similarly, $r_{\rm DP}$ characterises the region over which direct radiation pressure can overcome the confining pressure. When $r_{\rm DP}<r_{\rm acc}\leq r_{\rm PH}$, photoheating alone satisfies this condition across the whole accretion region. If both radii remain below $r_{\rm acc}$, neither mechanism alone does so. The two mechanisms also operate on different time-scales. Direct radiation pressure transfers momentum upon absorption, while photoheated gas expands over an ionized-gas sound-crossing time. Direct radiation pressure may therefore be particularly effective at rapidly suppressing accretion once $r_{\rm DP}$ reaches $r_{\rm acc}$.

\begin{figure*}
    \centering
    \includegraphics[width=\textwidth]{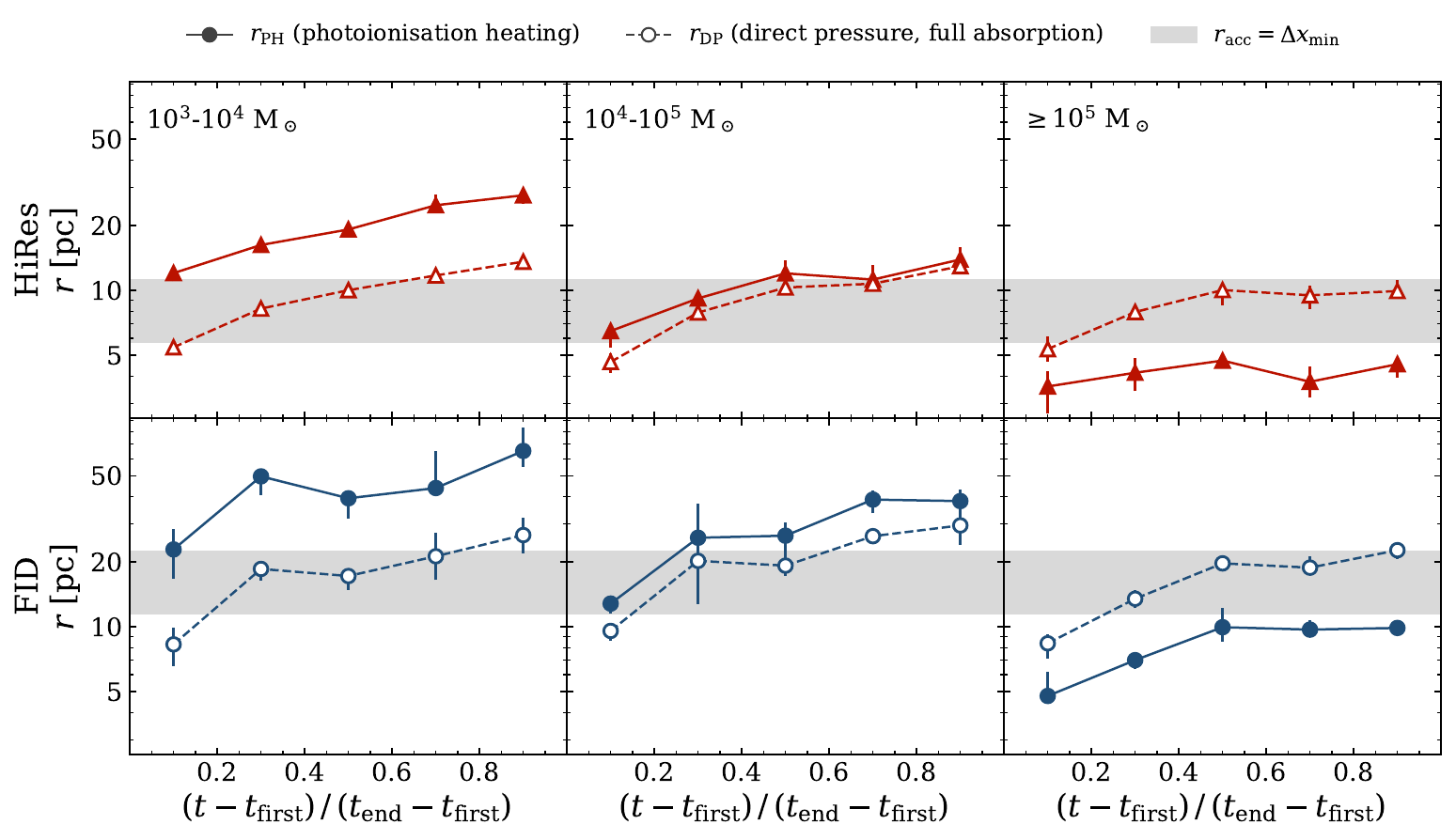}
    \caption{Pressure-balance radii throughout the first accretion episode for \hires (top) and \fid (bottom), grouped by sink particle birth mass (columns). Filled symbols and solid lines show photoionization heating radii, and open symbols and dashed lines direct radiation pressure radii under the assumption of full absorption. Both use the sum of gravitational and inward ram pressure measured within the accretion volume at snapshots as an estimate of confinement pressure. The measurements require that a simulation snapshot falls during the first accretion episode, which happens for 570 \hires and 197 \fid sink particles. Points require at least three sink particles and are medians in bins of the elapsed fraction of the formation episode duration, each of width 0.2, with each sink particle weighted equally. Error bars indicate 16th--84th percentile sink particle-bootstrap intervals. Lines join the ensemble medians. The grey band spans the physical accretion radii, $r_{\rm acc}=\Delta x_{\rm min}$, in each run, across the whole redshift range.}
    \alttext{Six panels compare photoheating and direct-radiation-pressure balance radii through cluster formation, with HiRes above, FID below, and increasing birth mass from left to right. Photoheating has the greater reach in the lowest-mass bin. Direct radiation pressure becomes larger in the highest-mass bin and approaches or exceeds the shaded accretion-radius band late in formation.}
    \label{fig:radii_phase}
\end{figure*}

Figure~\ref{fig:radii_phase} shows a mass-dependent change in the expected shutdown mechanism. Below $10^4\,{\rm M_\odot}$, photoheating can typically pressurise the accretion region early and retains the greater reach as direct radiation pressure strengthens. At $10^4$--$10^5\,{\rm M_\odot}$, neither channel typically reaches the accretion scale early, but both do later. Above $10^5\,{\rm M_\odot}$, the photoheating radius remains below the accretion scale, while direct radiation pressure typically reaches it late in the formation stage. This favours direct radiation pressure as the accretion shutdown mechanism for the most massive sink particles. The evolution throughout the cluster formation phase mainly shows the accretion scale being crossed while the ordering of the pressure-balance radii persists. The coarse snapshots sample an ensemble of star clusters at different fractions of their formation time, so they do not track individual accretion shutdowns or separate the causal contributions of the two radiative processes to these shutdowns. The measurements presented in Figure~\ref{fig:radii_phase} only include 570 of 1469 \hires clusters and 197 of 385 \fid clusters, as they require that at least one simulation snapshot falls during the first accretion episode. This requirement favours longer-lived accretion episodes when the formation time of a cluster is shorter than the snapshot spacing.

For a cold uniform medium, the same argument links the pressurisation to both stellar luminosity and ambient gas density (see Appendix~\ref{app:radii} for a derivation of the scalings). Assuming a density of $n_0=10^3\,{\rm cm^{-3}}$ and a temperature of $T_0=100\,{\rm K}$ for the cold ambient medium, our zero-age \textsc{bpass} output gives photo-heating and direct radiation pressure threshold masses of about $8$--$470$ and $330$--$5200\,{\rm M_\odot}$ respectively across cell sizes of 5.7--22.6\,pc. These values are low because this benchmark assumes that the expansion is only opposed by the thermal pressure of the cold ambient gas. In fact, according to this calculation, photoionization heating on its own should be enough to prevent any cluster more massive than $m_{\rm p}=10^3\,{\rm M_\odot}$ from forming. However, gravity and radial inflow provide a much stronger confinement of the gas around our sink particles, so these reference masses do not predict a shutdown of their accretion. This illustrates why the thermal pressure of the cold and dense ambient medium alone cannot account for the confinement and supports our approximation of neglecting it when estimating $P_{\rm conf}$. Note that values of cluster masses below $m_{\rm p}=10^3\,{\rm M_\odot}$ also extrapolate our assumption of properly sampling a stellar population beyond its domain of validity.

A further concern is that, unlike in our unresolved clusters, well-resolved cloud cores allow radiation to escape through low-density gas channels, and that opposing momentum input from spatially distributed stellar sources within a forming cluster reduces the net radial gas momentum about the cluster centre \citep{raskuttiNumericalSimulationsTurbulent2016,kimModelingUVRadiation2018}. Nevertheless, compact clusters such as the ones formed in our simulations can pass through an early stage dominated by direct radiation pressure before photo-heating takes over as gas is dispersed \citep{hanImpactRadiationFeedback2022a}. We also chose to neglect infrared trapping, which would add radiation pressure through dust reprocessing. This is because a large boost is unlikely at our low metallicities, consistent with the small infrared optical depths measured in the simulations of metal-poor cloud collisions of \citet{hanImpactRadiationFeedback2022a}. Appendix~\ref{app:cell} (more specifically Fig.~\ref{fig:cell}) charts the time evolution of the luminous output of our clusters and the physical state of the gas in their host cell throughout cluster formation in more detail.

\subsection{From the cloud to the cluster mass function}
\label{sec:chain}
\label{sec:disc_selection}

We test the connection between gas supply and the ICMF by assigning each main-galaxy sink particle of each run a final mass from equation~(\ref{eq:mstar_pressure}), using its measured initial accretion rate and drawing a value of $K$ from the distribution of couplings obtained by fitting the very same sink particle data. For the \norad run, we use its measured initial accretion rates and draw $K$ from the fitted distribution in \fid to predict a final mass. Figure~\ref{fig:chain} compares these mass functions with the measured ICMFs and natal-cloud distributions. Because the couplings are calibrated on the same simulations, this tests descriptive consistency rather than independent prediction.

This procedure reproduces the numbers of clusters with masses above $10^5\,{\rm M_\odot}$ measured in the simulations to within about 20~per~cent in \hires and \fid, but underestimates cluster counts at the high-mass end, particularly in \hires. This is because rapidly fed sink particles can receive fresh gas late in their growth, producing more scatter than the fitted distribution of $K$ captures (see Appendix~\ref{app:chain}). Applying \fid's cluster growth model to \norad reduces its number of clusters with masses above $10^5\,{\rm M_\odot}$ from 210 to 80. At fixed natal-cloud mass, \norad's initial accretion rates on to sink particles are slightly lower than \fid's, so its larger cluster masses arise from longer accretion episodes.

\begin{figure*}
    \centering
    \includegraphics[width=\textwidth]{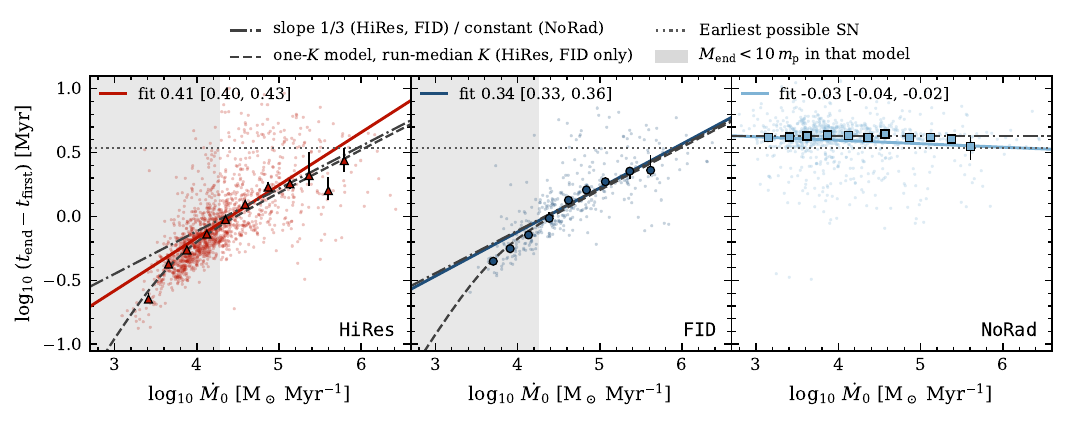}
    \caption{Accretion duration from $t_{\rm first}$ to $t_{\rm end}$, against the initial accretion rate measured at $t_{\rm first}$, for the birth sample in \hires, \fid and \norad. Points are individual sinks. Filled symbols are medians in 0.25-dex bins containing at least ten sink particles, with 16th--84th percentile sink particle-bootstrap intervals. Solid lines are full-range log--log regressions. Dash-dotted lines show a reference slope of value one third in the radiation runs and a constant duration in \norad. Dashed curves in \hires and \fid show the prediction of the $p=1/2$ model with the fitted median couplings $K=281$ and $260\,{\rm M_\odot^{1/2}\,Myr^{-2}}$, respectively. Shading marks rates for which clusters in that model end up with masses below ten star-particle masses. Dotted lines indicate the earliest possible supernova, $\sim3.4\,{\rm Myr}$ after $t_{\rm first}$.}
    \alttext{Three panels show accretion duration versus initial accretion rate. In HiRes and FID, durations generally increase with rate, with curvature and flattening relative to straight log-log fits. In NoRad, durations remain nearly constant near the earliest-supernova timescale. Points, binned medians and model curves show the scatter and the contrasting trends.}
    \label{fig:duration_rate}
\end{figure*}

\begin{figure*}
    \includegraphics[width=\textwidth]{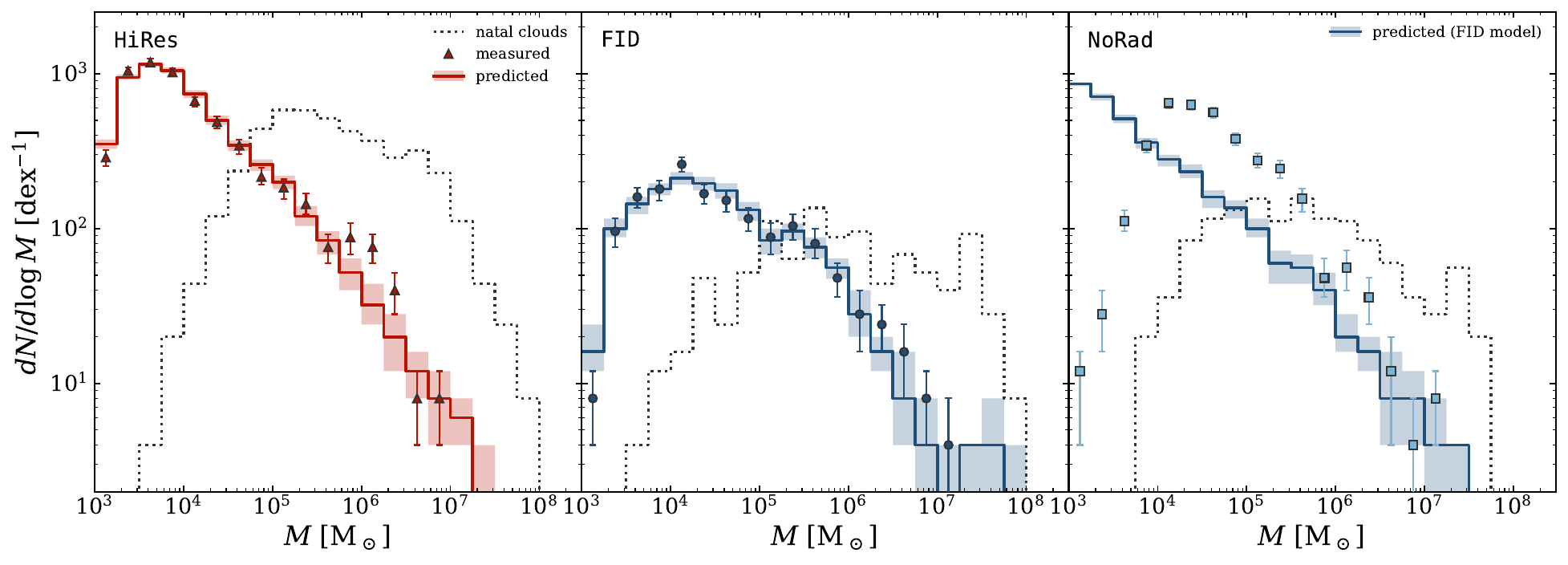}
    \vspace{-5mm}
    \caption{From natal clouds to the ICMF at $n_{\rm H}>100\,{\rm cm^{-3}}$ in \hires, \fid and \norad (left to right), as total number of objects per dex in each run. The dotted histogram shows natal-cloud detections matched to the birth sample at every catalogued output. Symbols represent the measured ICMF with 16th--84th percentile bootstrap error bars. The solid histogram and band give the median and 16th--84th percentiles over 300 realizations of equation~(\ref{eq:mstar_pressure}) with $p=1/2$, using the initial accretion rate of each sink particle and drawing the coupling $K$ from the individual-track fits of the run it belongs to. \norad uses its own accretion rates but \fid's couplings and is thus not fitted.}

    \alttext{Three panels compare natal-cloud mass distributions, measured cluster birth mass functions and growth-model predictions for HiRes, FID and NoRad. The model broadly follows the radiation-run cluster counts but underpredicts the high-mass tail. Applying FID couplings to NoRad predicts fewer massive clusters than measured. Bands show variation among model realizations.}
    \label{fig:chain}
    \vspace{-3mm}
\end{figure*}

Figure~\ref{fig:duration_rate} shows duration against initial accretion rate. In the radiation runs, rapidly fed sinks also accrete for longer. Full-range log--log slopes are $0.41\pm0.02$ in \hires and $0.35\pm0.02$ in \fid, but the relation is curved. The one-third asymptotic prediction for the $p=1/2$ empirical model is most useful over $10^4$--$10^5\,{\rm M_\odot\,Myr^{-1}}$. The model curve bends below this scaling at low rates because the finite mass $m_{\rm p}$ of the first stellar particle cannot be neglected in equation~(\ref{eq:mstar_pressure}). When little mass is subsequently accreted, the first particle already supplies much of the feedback, shortening the remaining growth. The discrete stellar mass scale therefore breaks the model's scale-free behaviour at low rates. The measured relation also flattens at high rates. In \norad the duration of the accretion episode is nearly independent of the initial rate and remains close to the first-supernova explosion time delay.

This difference changes the ICMF slope. If ${\rm d}N/{\rm d}\dot M_0\propto\dot M_0^{\beta_{\dot M}}$ and $M_\star\propto\dot M_0^\kappa$, then $\beta_{\rm pl}=(\beta_{\dot M}+1)/\kappa-1$. Note that this relation neglects any scatter in cluster growth. Above the adopted mass limits, the fitted mass exponents are $\kappa=1.29$ and 1.28 in \hires and \fid, compared with 0.83 in \norad. With stellar radiation, faster feeding is associated with longer growth, so the final mass rises more steeply than linearly with the initial rate and the ICMF becomes shallower. In \norad, the nearly rate-independent duration is generally limited by the first-supernova delay. Faster feeding then increases mass without also extending the available growth time, giving a mapping that is closer to linear, though still sublinear, thus leading to a steeper ICMF. Radiation changes both the duration of accretion and its dependence on the gas supply. This explains how it can shorten cluster formation overall while producing a shallower mass function.
The slopes implied by the empirical growth model agree with the measured power-law slopes to within 0.04 in the radiation runs (see Appendix~\ref{app:chain}). The larger number of low-mass clouds forming sink particles in \norad also contributes to its steeper ICMF (Section~\ref{sec:clouds_mf}), so growth histories and the cloud population are not cleanly separated.

\label{sec:truncation}
The growth model does not establish what sets the ICMF upper mass cut-off. Its fitted value is sensitive to the behaviour of a small number of massive clusters. Some rapidly fed sinks stop earlier than predicted and have weaker inflows on larger scales, suggesting that external gas supply also matters at the top end of the cluster mass function. The host galaxy may therefore contribute to setting the upper mass scale of the ICMF, although our fitted mass cut-offs are not accurate enough to rule out an impact from radiative feedback. Observed truncation masses rise with star formation rate surface density \citep{johnsonPanchromaticHubbleAndromeda2017,adamoStarClusterFormation2020}, while analytic models link them to gas surface density, shear and the time available for collapse before feedback \citep{reina-camposUnifiedModelMaximum2017a}. Testing these dependencies requires a larger sample of galaxies with different masses and environments.

%==============================================================================
\subsection{Comparison with other simulations and observed cluster populations}
\label{sec:disc_sims}

In an isolated dwarf, \citet{anderssonPresupernovaFeedbackSets2024a} find that early feedback, especially ionizing radiation, suppresses massive clusters and steepens their mass function. Our comparison between \fid and \norad instead shows that the ICMF becomes shallower with stellar radiation. Their clusters assemble from distributed $100\,{\rm M_\odot}$ star particles formed at a prescribed efficiency per free-fall time, and many low-mass clusters continue forming beyond the first possible supernova. Our sink particles accrete surrounding gas according to how much of this gas flows into them and concentrate all their radiation emission in one spatial point. The differing growth histories and source geometry may contribute to the contrast, alongside the galactic environment. Our galaxy also forms substantially more stellar mass at fixed halo mass than the empirical relations shown in Fig.~\ref{fig:smhm}. Its shallow ICMF prediction should therefore not be treated as that of the typical galaxy at cosmic dawn.

\citet{liStarClusterFormation2017} also obtain an emergent Schechter-like ICMF in cosmological simulations, with a truncation mass at the top end that increases with SFR. Their cluster particles grow within spheres of fixed 10\,pc diameter at a prescribed efficiency per free-fall time until feedback interrupts the gas supply. Our growth rate instead follows the inflow through the accretion zone. In both approaches, the mass function emerges from cluster growth, but its relation to the available gas depends on the accretion prescription.

Our truncation masses fall within the range seen in nearby mergers. The HiPEEC sample spans $M_{\rm c}\sim10^6$ to a few $10^7\,{\rm M_\odot}$ \citep{adamoStarClusterFormation2020}, whereas nearby spirals typically truncate near $10^5\,{\rm M_\odot}$ \citep{gielesLuminosityFunctionYoung2006,adamoProbingRoleGalactic2015}. The generally steeper ICMFs of these galaxies may reflect different environments or feedback channels omitted in our modelling, such as cosmic rays.  Growing lensed samples will also test the slope of the ICMF below the truncation mass at high redshift \citep{claeyssensFirstGLIMPSEStar2026} in the near future.

%==============================================================================

\subsection{Comparing with \textit{JWST} lensed clusters}
\label{sec:disc_obs}

Our brightest clusters reach the UV luminosities of the high-redshift \textit{JWST} lensed sample at a given nominal SFR averaged over $\sim$100~Myr for the host galaxy (Section~\ref{sec:brightest}). The similar offsets of the three runs despite their different ICMFs show that this statistic does not uniquely recover the birth mass function. Sampling, burstiness and stellar populations gained after the first episode also affect the brightest source.

The comparison is sensitive to how clusters and host SFRs are measured. Magnification affects both luminosity and SFR, and lens-model systematics are excluded from our quoted offsets. Unresolved neighbours can blend into brighter sources, while detection limits favour hosts with bright clusters \citep{pfefferComparingEMOSAICSPredictions2025}. Our sink particles do not have resolved sizes, so we cannot account for observational size selection. The host SFR tracers also average star formation over different periods. UV light responds to recent star formation history rather than to a fixed 100\,Myr window \citep{floresvelazquezTimescalesProbedStar2021}, while recombination lines probe younger stars and SED-based SFRs depend on the detailed assumptions made about the star formation history \citep{choeUncertaintiesHighzGalaxy2026}.

Our simulated magnitudes also account for stellar light alone, whereas young observed clusters can have appreciable nebular continuum at 1500\,\AA\ \citep{katz21BalmerJump2025}. Assuming that all of a sink particle's ionizing photons produce nebular emission at $10^4\,{\rm K}$ gives a brightening of a few tenths of a magnitude in the first few Myr, falling below a tenth after 10\,Myr. Escape and dust absorption reduce this upper limit. Correcting the youngest observed sources for nebular light would make their stellar magnitudes fainter and move them towards the $\alpha=-2$ reference power law ICMF.

Monte Carlo predictions derived from specific ICMFs assume constant star formation with all stars forming in clusters. A lower cluster formation efficiency would reduce the expected luminosity of the brightest cluster, and burstiness in the SFR changes the sampled population. With one simulated halo and ten observed lensed systems, the comparison establishes that our model can produce sufficiently bright clusters. It cannot discriminate between different ICMFs or feedback prescriptions.

%==============================================================================

\subsection{Limitations and excluded physics}
\label{sec:limits}

\subsubsection{Pre-supernova feedback}

We include photoionization heating and direct radiation pressure as pre-supernova feedback, but omit stellar winds \citep{geenGeometryDynamicalRole2020,lancasterEfficientlyCooledStellar2021}, protostellar jets \citep{federrathModelingJetOutflow2014}, infrared trapping \citep{menonInfraredRadiationFeedback2022} and Lyman-$\alpha$ pressure. The latter can matter in metal-poor gas, depending on its structure and kinematics \citep{kimmImpactLymanAlpha2018,nebrinLymanaFeedbackPrevails2025}. The geometry and resolution dependence of the direct radiation pressure estimate is discussed in Section~\ref{sec:shutdown}.

Radiation has a stronger effect than stellar winds on cluster formation in the dwarf simulations of \citet{anderssonPresupernovaFeedbackSets2024a}, and \citet{quinalhaStellarFeedbackEffects2026} find that stellar winds also have little additional effect on their cloud mass function. These comparisons do not determine the importance of these omitted feedback channels in our cosmological galaxy simulations, where they could alter both the cloud population and the local accretion.

\subsubsection{Sink particles and a single halo}

The cluster seeding threshold depends on gas temperature and resolution through $c_{\rm s}^2/\Delta x_{\rm min}^2$, affecting where sink particles form. Their accretion zone spans one cell radius. The flux scheme reproduces accretion on to a collapsing isothermal sphere and is not strongly sensitive to the chosen accretion radius \citep{kangImpactStarFormation2025}, but resolved gas structure still affects local growth and feedback. While sink particles that end up with masses below $10^4\,{\rm M_\odot}$ become field stars after formation and remain in our birth-property analyses, mergers between sink particles occur when they become bound and their accretion zones overlap, potentially restoring accretion on to clusters that have already formed. This numerical condition need not imply a physical cluster merger. The accretion rate is usually already zero before our numerical cluster formation shutdown criterion is met (Section~\ref{subsec:sink_star}), but disabling subsequent accretion also prevents later growth that might otherwise resume without a merger.

A sink particle is a point mass which carries various stellar populations, so we do not follow internal cluster dynamics, gas-expulsion losses or tidal disruption \citep{liDisruptionGiantMolecular2019,brownTestingFeedbackStar2022}. Our birth masses measure stars formed rather than bound mass. Extending the simulations beyond this single halo and to lower redshift will test environmental trends and connect these birth populations to surviving clusters.

\section{Conclusions}
\label{sec:conclusions}

We have studied star cluster formation in cosmological zoom-in radiative hydrodynamics simulations of a halo reaching $\sim10^{10}\,{\rm M_\odot}$ at $z=6$. Sink particles, representing forming star clusters, accrete gas and retain the stellar populations they form without recourse to a prescribed star formation efficiency per free-fall time or fixed accretion duration, so the ICMF naturally emerges from gas supply and stellar feedback in the simulations. We compare simulations of the same halo which include stellar radiation at two different spatial resolutions and a run without stellar radiation at the fiducial resolution. All three include supernova feedback. Our main findings are as follows.

\begin{itemize}
    \item Schechter fits to the ICMF give truncation masses of about $6\times10^6\,{\rm M_\odot}$, with overlapping posterior distributions and substantial uncertainty. At the fiducial resolution, stellar radiation makes the low-mass power-law slope of the ICMF shallower, from $-1.60$ to $-1.34$. The higher-resolution run, \hires, yields an intermediate power-law slope of $-1.54$ and produces about 1.6 times as many clusters above $10^5\,{\rm M_\odot}$ as its lower-resolution counterpart \fid. The low-mass slope of the ICMF and the abundance of star clusters therefore remain sensitive to resolution.

    \item Stellar radiation reduces the abundance of low-mass clouds, most strongly at low density, and raises the sink particle-seeding density threshold by heating the surrounding ISM gas. The small number of massive clusters (above the truncation mass) produced in the simulations prevents us from reaching robust conclusions regarding the effect of radiative feedback at the upper end of the ICMF.

    \item In each radiative run, about 92--94~per~cent of the analysed main-galaxy clusters stop accreting before the earliest possible supernova from their own stellar populations goes off, with median accretion durations of about a Myr. On average, more rapidly fed clusters accrete for longer, making their final masses rise faster than linearly with the initial gas accretion rate. Without stellar radiation, accretion typically persists beyond the minimum supernova time delay, with a duration only weakly dependent on the initial accretion rate, and the mapping from initial gas accretion rate to final stellar mass is closer to linear. Under the adopted assumptions about pressure confinement and full absorption of stellar luminosity, pressure-balance estimates are consistent with photoionization heating limiting low-mass cluster growth and direct radiation pressure becoming important late in the growth of the most massive clusters. An empirical braking growth law describes much of the mapping from accretion rate to cluster mass in the simulations, but does not fully capture the high-mass tail behaviour or establish the origin of the mass truncation.

    \item Our brightest simulated clusters reach the UV luminosities of observed \textit{JWST} lensed clusters hosted by galaxies with similar SFRs. The comparable offsets in cluster brightness relative to fixed ICMF predictions across our runs demonstrate that this statistic alone cannot distinguish the underlying ICMFs. It is affected by sampling, burstiness of star formation histories and potential mergers between clusters.
\end{itemize}

Despite the limitations of our analysis -- omitted stellar feedback channels, cluster birth masses that measure stars formed rather than the fraction remaining bound, and a galaxy sample too limited to test the environmental dependence of the ICMF -- our simulations reveal a key link between stellar radiation and the shape of the ICMF at cosmic dawn. Following cluster dynamics and mass loss should allow us to connect these early populations to the fossil record that has survived to the present day. We plan to improve our modelling to address these issues in future work.

\section*{Acknowledgements}

FST acknowledges support from the Science and Technology Facilities Council (STFC) for a PhD studentship and from Yonsei University for hosting and support during a research visit. TK and CK were supported by the National Research Foundation of Korea (RS-2022-NR070872 and RS-2025-00516961). This work used the DiRAC Data Intensive service (DIaL3) at the University of Leicester, managed by the University of Leicester Research Computing Service on behalf of the STFC DiRAC HPC Facility (www.dirac.ac.uk). The DiRAC service at Leicester was funded by BEIS, UKRI and STFC capital funding and STFC operations grants. DiRAC is part of the UKRI Digital Research Infrastructure. This work used the DiRAC Memory Intensive service (COSMA8) at Durham University, managed by the Institute for Computational Cosmology on behalf of the STFC DiRAC HPC Facility (www.dirac.ac.uk). The DiRAC service at Durham was funded by BEIS, UKRI and STFC capital funding, Durham University and STFC operations grants. DiRAC is part of the UKRI Digital Research Infrastructure. The authors thank Jonathan Patterson for maintaining the Glamdring Cluster hosted by the University of Oxford, where part of the data processing was performed. We thank the Astrophysics Data Service and the arXiv preprint repository for providing open-source software and services that were used extensively in this work.

%%%%%%%%%%%%%%%%%%%%%%%%%%%%%%%%%%%%%%%%%%%%%%%%%%
\section*{Data Availability}
The data underlying this article will be shared on reasonable request to the corresponding author.


\bibliographystyle{mnras}
\bibliography{first_paper} % if your bibtex file is called example.bib

%%%%%%%%%%%%%%%%% APPENDICES %%%%%%%%%%%%%%%%%%%%%

\appendix
\section{The observed comparison sample}
\label{app:obs}

Table~\ref{tab:obs} lists the lensed systems used in Section~\ref{subsec:obs}. We convert the published photometry of each brightest compact source to an intrinsic rest-frame magnitude at 1500\,\AA, correcting for magnification where required, for dust using the source's measured reddening and the \citet{calzettiDustContentOpacity2000} law, and for rest wavelength using the measured UV slope. The rest wavelength correction is at most 0.4\,mag and the dust correction at most 1.3\,mag. Quoted errors include photometric and statistical magnification uncertainties and the wavelength correction, but exclude dust uncertainty and lens-model systematics. The latter span 0.3--2.4\,mag and are one-sided towards fainter magnitudes for the Sunrise and Cosmic Gems arcs. The Sunrise magnification is a lower limit, while the Cosmic Gems value we use is the lowest of three lens models. Magnification affects both cluster luminosity and host SFR estimates.

The $\sim$100~Myr averaged SFRs come from the published UV or SED estimates. For the $\sim$10~Myr SFR averages we use published recombination-line fluxes and magnifications with the \citet{kennicuttStarFormationMilky2012a} H$\alpha$ calibration for four systems, and the published recent SED-based SFRs otherwise. Incomplete host coverage gives lower SFR limits, including the Cosmic Spear's single-clump estimate. Host masses and cosmologies are retained as published.

\begin{table*}
\centering
\begin{threeparttable}
\caption{Observed lensed systems and their brightest compact sources. Columns give redshift, intrinsic rest-frame UV magnitude, source age, host stellar mass and SFR estimates on the two averaged time-scales of 10 and 100 Myr. Magnitudes are corrected for lensing, dust and rest wavelength as described in the text. Masses and SFRs are de-lensed, with SFRs in ${\rm M_\odot\,yr^{-1}}$.}
\label{tab:obs}
\begin{tabular}{l l c c c c c c l}
\hline\hline
System & ID & $z$ & $M_{1500}$ & Age [Myr] & $\log M_{\star,\rm host}$ & SFR$_{100}$ & SFR$_{10}$ & References \\
\hline
RCS0224 & E$_1$ & 4.88 & $-18.94\pm0.39$ & 1--20$^{e}$ & -- & -- & $8.2\pm1.4$ & 1, 2 \\
Sunrise arc & 6b & 6.0$^{a}$ & $-17.10\pm0.39$ & $4^{+2}_{-3}$ & 8.5--9.3 & -- & $6.5\pm3.5^{b}$ & 3 \\
Cosmic Grapes & clump 1 & 6.072 & $-18.86\pm0.17$ & $6.9\pm2.7^{f}$ & 8.65 & 2.6 & $13.8^{+3.8}_{-0.9}\,^{c}$ & 4 \\
MACS0416 & D1 & 6.144 & $-16.26\pm0.20$ & $12^{+0}_{-1}$ & 7.5$^{d}$ & $0.35$ & $0.45^{+0.06}_{-0.04}\,^{c}$ & 5 \\
MACS0416 & T1 & 6.145 & $-15.03\pm0.25$ & $2^{+6}_{-0}$ & -- & $0.12\pm0.01$ & $0.49^{+0.13}_{-0.10}\,^{c}$ & 5 \\
MACS0416 & UT1 & 6.145 & $-14.73\pm0.30$ & $14^{+57}_{-0}$ & -- & $0.03\pm0.01$ & -- & 5 \\
Cosmic Spear & 3abc & 6.2 & $-18.00\pm0.22$ & $5^{+2}_{-1}$ & 8.64 & -- & $>12^{+7}_{-5}$ & 6 \\
Firefly Sparkle & brightest & 8.296 & $-17.43\pm0.25$ & -- & 6.8 & -- & $0.81^{+0.27}_{-0.19}\,^{c}$ & 7 \\
Cosmic Gems & A.1 & 9.625 & $-16.02\pm0.62$ & $9.2^{+13.5}_{-3.2}$ & $>$7.4--7.8 & $>0.33$ & -- & 8, 9, 10 \\
Misty Moons & ID1 A.1 & 11.5$^{a}$ & $-17.31\pm0.31$ & $42^{+27}_{-19}$ & 7.57 & $0.32$ & $0.60\pm0.12$ & 11 \\
\hline
\end{tabular}
\begin{tablenotes}
\item $^{a}$ Photometric redshift. $^{b}$ Midpoint of the published SED range (3--10). $^{c}$ Derived here from line fluxes or $\xi_{\rm ion}$ and magnifications. $^{d}$ The full MACS0416 system. $^{e}$ Range of acceptable age fits. $^{f}$ Compiled by \citet{pfefferComparingEMOSAICSPredictions2025}. Ages depend on the assumed star formation history. $>$ denotes incomplete host coverage. Host masses retain the source IMF (Chabrier for the Sunrise arc and Misty Moons, Kroupa otherwise), with the Sunrise range spanning two fits.
\item References are (1) \citet{messaPropertiesBrightestYoung2024}, (2) \citet{swinbankResolvedSpectroscopyGravitationally2007}, (3) \citet{vanzellaJWSTNIRCamProbes2023}, (4) \citet{fujimotoPrimordialRotatingDisk2025}, (5) \citet{messaAnatomyZ6Lymana2025}, (6) \citet{abdurroufSpatiallyResolvedPhysical2025}, (7) \citet{mowlaFormationLowmassGalaxy2024}, (8) \citet{adamoBoundStarClusters2024a}, (9) \citet{bradleyUnveilingCosmicGems2025}, (10) \citet{messaJWSTSpectroscopicConfirmation2026}, (11) \citet{nakaneVENUSStronglyLensed2025}. Photometry for the Cosmic Grapes, MACS0416 and the Firefly Sparkle as compiled by \citet{pfefferComparingEMOSAICSPredictions2025}.
\end{tablenotes}
\end{threeparttable}
\end{table*}

\section{Fitting the initial cluster mass function}
\label{app:icmf}
\label{app:icmf_mmin}

\begin{figure*}
    \includegraphics[width=\textwidth]{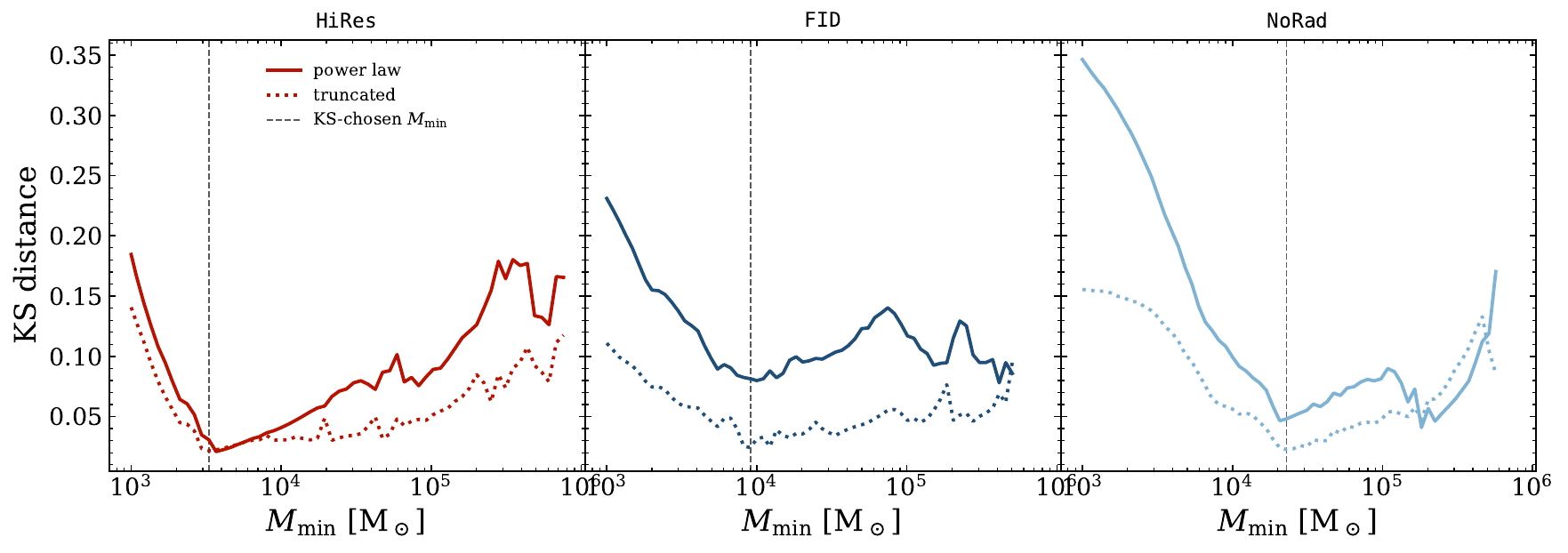}
    \vspace{-5mm}
    \caption{Kolmogorov--Smirnov distance for power-law (solid) and Schechter (dotted) fits as the lower mass limit varies. Panels show the birth sample defined in the text for \hires, \fid and \norad respectively. Dashed lines mark the adopted lower mass limits, chosen as the lowest scanned value within 3~per~cent of the minimum Schechter distance, as used in Figure~\ref{fig:icmf}.}
    \alttext{Three panels plot Kolmogorov--Smirnov distance against the lower fitting mass for HiRes, FID and NoRad. Schechter curves generally lie below power-law curves, particularly in FID and NoRad. Both distances vary non-monotonically. Vertical dashed lines mark the adopted lower limits near low-distance regions of the Schechter scans.}
    \label{fig:A3}
    \vspace{-3mm}
\end{figure*}

Numerical effects may alter the low-mass ICMF, so we restrict the fits to masses above $M_{\rm min}$. We scan the lower mass fitting limit and pick the lowest $M_{\rm min}$ whose Kolmogorov--Smirnov distance is within 3~per~cent of the minimum for maximum-likelihood Schechter fits (Fig.~\ref{fig:A3}). Here $M_{\rm min}$ is an estimated fitting threshold, rather than a prescribed numerical completeness limit. The adopted mass limits are $3.3\times10^3$, $9.0\times10^3$ and $2.3\times10^4\,{\rm M_\odot}$ in \hires, \fid and \norad respectively. Both the power law and the Schechter function are fitted above this Schechter-selected limit in each run.

We assess goodness of fit with 10\,000 synthetic catalogues for each law, adapting the method of \citet{clausetPowerlawDistributionsEmpirical2009}. Each catalogue has the size of the full sample. Each mass is drawn from the maximum-likelihood law above the adopted $M_{\rm min}$ with probability equal to the measured fraction of clusters above that limit, while masses below the limit are resampled from the measured masses. We then repeat the lower-limit scan and parameter fitting on each catalogue, using the same Schechter-based 3~per~cent rule relative to that catalogue's own minimum. The $p$-value is the fraction of synthetic catalogues whose Kolmogorov--Smirnov distance for the law being tested is at least as large as the measured one. This accounts for the selection of the fitting range.

Table~\ref{tab:icmf_gof} presents the results. At our adopted 5~per~cent rejection threshold, the Schechter fits are not rejected in any run, while pure power laws are rejected in all three. These goodness-of-fit tests do not directly measure the relative preference between the two fitted laws.

\noindent\begin{minipage}{\columnwidth}
    \centering
    \captionsetup{hypcap=false,labelfont=bf,labelsep=period}
    \captionof{table}{ICMF goodness-of-fit $p$-values from synthetic catalogues analysed with the same lower-limit selection and fitting procedure as the simulation data.}
    \label{tab:icmf_gof}
    \begin{tabular}{lcc}
        \hline\hline
        Run & Schechter & Power law \\
        \hline
        \hires & 0.072 & 0.015 \\
        \fid & 0.847 & 0.0054 \\
        \norad & 0.496 & 0.013 \\
        \hline
    \end{tabular}
\end{minipage}\par\medskip

\begin{figure*}
    \includegraphics[width=\textwidth]{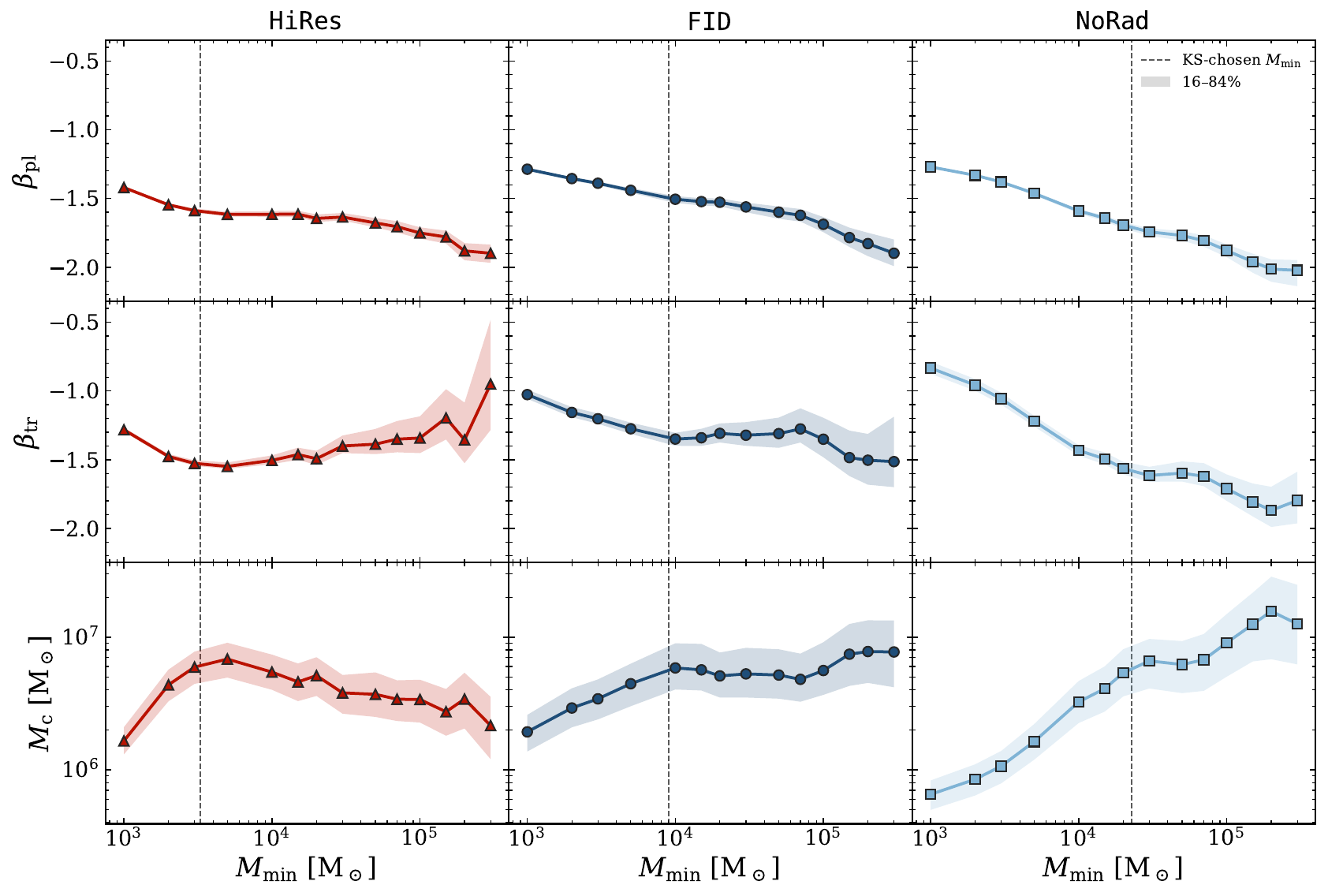}
    \vspace{-5mm}
    \caption{Dependence of the ICMF fits on the lower mass limit for main-galaxy clusters in \hires, \fid and \norad. Rows show the power-law slope, Schechter slope and truncation mass for fourteen different mass limits between $10^3$ and $3\times10^5\,{\rm M_\odot}$. Dashed vertical lines mark the adopted lower mass limits from Figure~\ref{fig:A3}. Bands show 16th--84th percentiles from cluster bootstrap resampling.}
    \alttext{A three-by-three grid shows fitted mass-function parameters versus lower fitting mass, with HiRes, FID and NoRad in columns. Power-law slopes steepen as the threshold rises. Schechter slopes and cut-off masses also vary, with the NoRad cut-off rising strongly. Shaded uncertainty bands and vertical adopted-threshold markers show the sensitivity of the inferred parameters.}
    \label{fig:A4}
    \vspace{-3mm}
\end{figure*}

Figure~\ref{fig:A4} tests fit sensitivity to $M_{\rm min}$. Power-law fits steepen as the lower mass limit rises, while Schechter fits trade slope against truncation mass. \fid remains the run which yields the shallowest slope above a common $10^4\,{\rm M_\odot}$ mass limit, but the slope contrast with \norad is smaller than when using separately selected mass limits. In \norad the fitted truncation mass varies by more than an order of magnitude across the scanned lower mass limits. The posterior intervals quoted in Section~\ref{sec:icmf} therefore describe uncertainty at the adopted mass limit, rather than this additional sensitivity.


\section{The growth model and the cloud-to-cluster mapping}
\label{app:chain}
\label{app:chain_params}

\begin{figure*}
    \includegraphics[width=\textwidth]{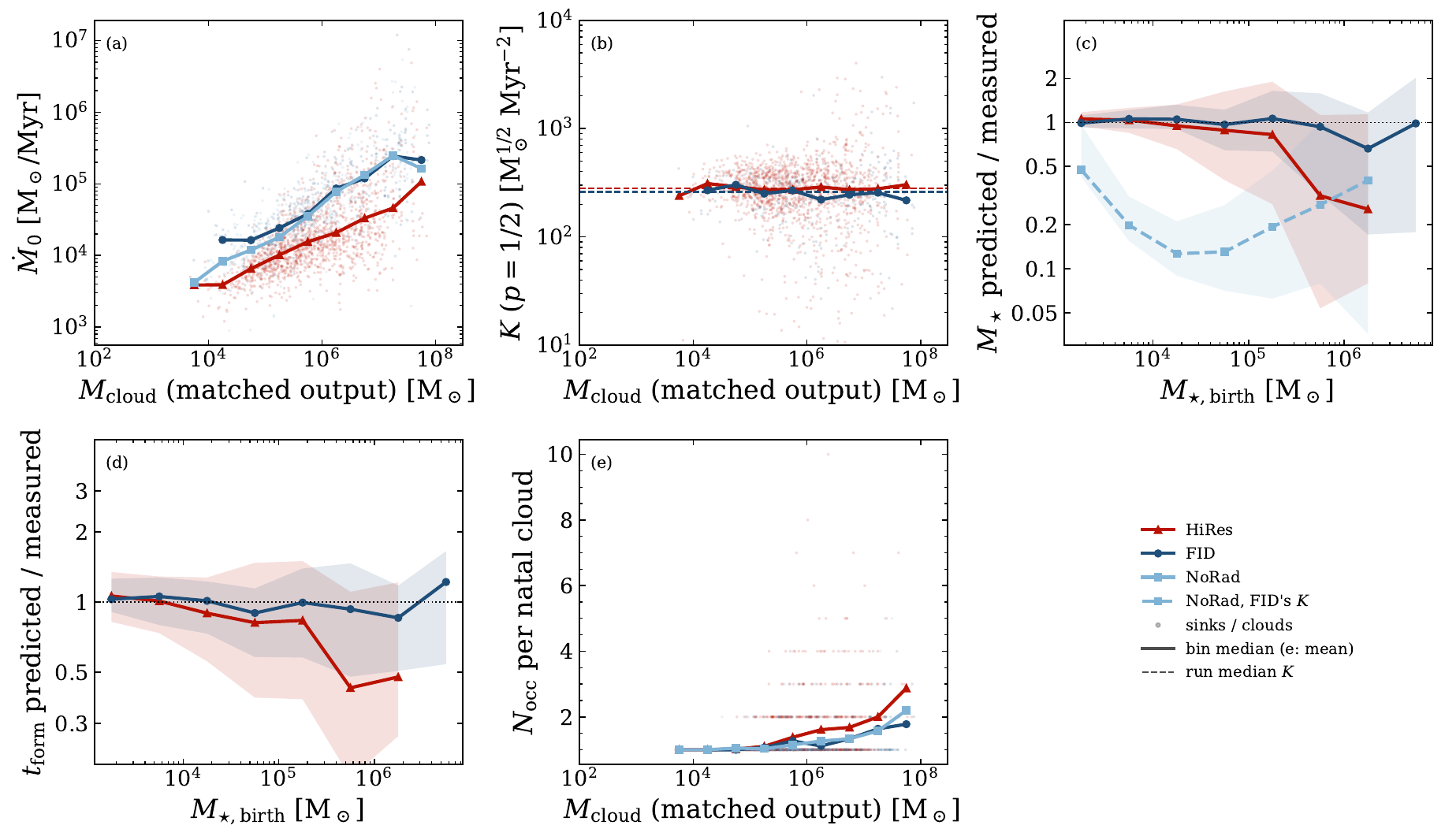}
    \vspace{-5mm}
    \caption{Diagnostics of the main-galaxy cloud-to-cluster mapping. Panels show (a) initial accretion rate against natal-cloud mass, (b) fitted coupling against cloud mass, (c) predicted-to-measured birth mass, (d) predicted-to-measured formation time and (e) sink count per cloud. Solid lines show medians in 0.5-dex bins, except for the mean in (c), with 16th--84th percentile bands in (c) and (d). Predictions use measured initial accretion rate for each sink and one median coupling per run, with \norad using \fid's coupling. Panel (b) contains 1459 and 306 sinks with natal-cloud matches in \hires and \fid, while (c) and (d) use the full mapping sample. The model is calibrated on these same simulations.}
    \alttext{Five panels show cloud-to-cluster diagnostics. Initial accretion rate increases with cloud mass, while fitted coupling is roughly constant. Predicted-to-measured mass and duration ratios are near unity over much of the radiation-run samples but decline for the most massive HiRes clusters; NoRad ratios are mostly lower. The final panel shows that more massive clouds host more cluster sinks.}
    \label{fig:A9}
    \vspace{-3mm}
\end{figure*}

Figure~\ref{fig:A9} supports the mapping discussed in Section~\ref{sec:chain}. Natal clouds are matched in every catalogued output, unlike the more widely spaced cloud census. Initial accretion rates rise with cloud mass, while the coupling $K$ remains roughly constant. The apparent high-mass deficit in panel (c) reflects binning by measured mass, which biases towards sink particles with unusually large growth. It disappears when binned by initial rate, although late gas supply broadens the mass distribution beyond the model scatter in $K$.

\label{app:chain_occupancy}
More massive clouds also host more sink particles, from one below a cloud mass of a few $10^4\,{\rm M_\odot}$ to two or three above $10^7\,{\rm M_\odot}$. We predict each sink particle's mass separately rather than dividing the cloud mass among all sink particles.
Above our adopted mass fitting limits for the ICMF, the initial accretion rate slopes are $-1.81$, $-1.68$ and $-1.69$ in \hires, \fid and \norad respectively. Combined with the measured exponents of the relation between initial accretion rate and final stellar mass, these imply ICMF slopes of $-1.63$, $-1.53$ and $-1.84$, within 0.04 of the measured power-law slopes in the radiation runs and 0.12 steeper in \norad. These power-law fits use separately selected lower mass limits for the 3 runs and describe the mapping sample rather than the sample used to perform the Schechter fits. The stellar mass already present at the time of initial accretion rate measurement prevents a direct conversion between the fitted duration--rate and mass--rate exponents.

\label{app:feedback_models}
We fit equation~(\ref{eq:pressure_model}) with $p=1/2$ to individual \hires and \fid accretion histories, fixing the initial stellar mass to $10^3\,{\rm M_\odot}$ and $\dot M_0$ to the accretion rate measured when the sink particle reaches this initial stellar mass. The single free parameter of the model, $K$, minimizes squared residuals in $\log_{10}\dot M$. The mass function mapping draws coupling values from these fits. The prediction bands in Fig.~\ref{fig:chain} represent variation between realizations that draw $K$ from this distribution, rather than uncertainty in the growth law. The predictions in Fig.~\ref{fig:A9} instead use one median $K$ per run, with bands showing the 16th--84th percentiles of the prediction ratios across sink particles. Alternative braking laws, including linear ($p=1$) mass dependence and saturation of thermal feedback, give similar mass predictions. Allowing the braking exponent, $p$, to vary introduces covariance between $K$ and the initial accretion rate without materially improving the description of the accretion histories. We therefore retain $p=1/2$ as a simple empirical description, without identifying a unique braking mechanism.

\section{The cell of a forming sink particle}
\label{app:cell}

\begin{figure*}
    \centering
    \includegraphics[width=\textwidth]{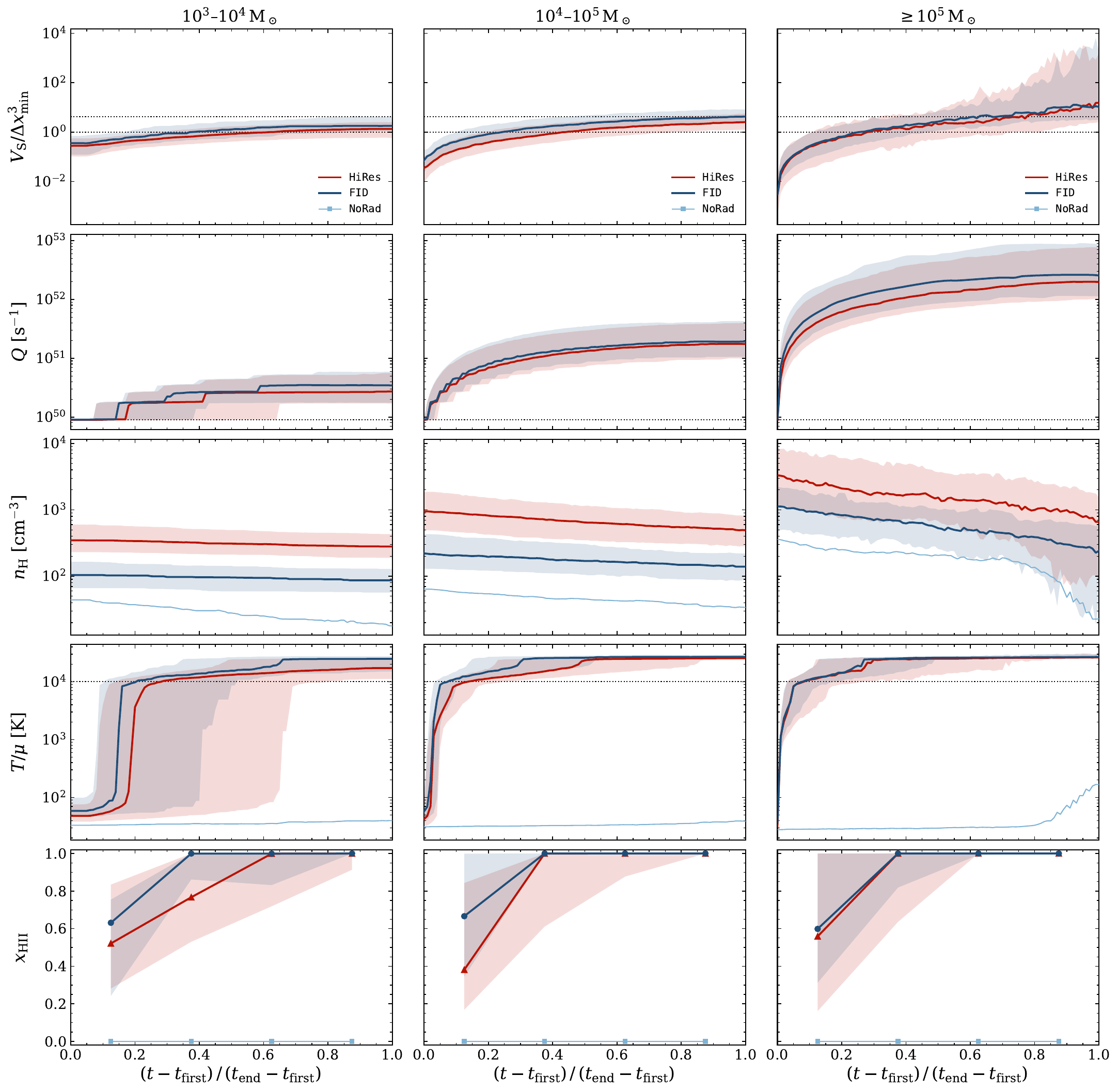}
    \vspace{-2mm}
    \caption{Cluster host cell evolution against elapsed fraction of the formation episode $(t-t_{\rm first})/(t_{\rm end}-t_{\rm first})$, grouped by sink particle birth mass (columns). Each diagnostic uses the birth sample described in the text, with the corresponding event or snapshot measurements. Rows show nominal Str\"omgren volume relative to cell volume, stellar ionizing photon rate, gas density, the temperature proxy $T/\mu$, where $\mu$ is the dimensionless mean molecular weight, and ionized hydrogen fraction. Solid lines and bands are medians and 16th--84th percentiles in \hires and \fid. \norad is shown without bands in the gas property rows only. Gas density and $T/\mu$ are measured at star-formation events (every time a cluster adds 10$^3$ M$_\odot$ to its mass). $Q$ and $V_{\rm S}$ are recomputed as stellar populations age. Ionised gas fractions use only four bins of the elapsed fraction of the formation episode, as they need to be measured from simulation snapshots spaced about 2\,Myr apart. Dotted lines mark one cell volume and $R_{\rm S}=\Delta x_{\rm min}$ in the top row, the zero-age output of a $10^3\,{\rm M_\odot}$ stellar population in the second, and $T/\mu=10^4\,{\rm K}$ in the fourth.}
    \alttext{A five-by-three grid shows cluster host-cell evolution, with increasing cluster birth mass in columns. Rows show ionized-volume ratio, ionizing photon rate, gas density, temperature proxy and ionized hydrogen fraction. In HiRes and FID, photon output rises, density falls, and gas rapidly becomes hot and ionized. NoRad gas remains much colder and predominantly neutral. Lines and bands show medians and scatter.}
    \label{fig:cell}
    \vspace{-3mm}
\end{figure*}

Figure~\ref{fig:cell} shows rapid heating of the gas cell hosting a forming cluster to $T/\mu=10^4\,{\rm K}$ in \hires and \fid, followed by growth of the nominal Str\"omgren volume as the ionizing photon rate rises and gas density falls. In \norad, the host cell remains cold and neutral, as expected. Solid curves describe the overall cluster population rather than individual accretion shutdown histories, and the nominal volume is a photon-budget estimate rather than a resolved ionization front.

\subsection{\texorpdfstring{Uniform cold-medium benchmark}{Uniform cold-medium benchmark}}
\label{app:radii}

For the uniform cold-medium benchmark discussed in Section~\ref{sec:shutdown}, we set $P_{\rm conf}=n_0k_{\rm B}T_0$ (where $n_0$ and $T_0$ are the density and temperature of the cold ambient medium respectively), $Q=qM_\star$ and $L_{\rm abs}=f_{\rm abs}\ell M_\star$ in equations~(\ref{eq:radius_photoheating}) and (\ref{eq:radius_direct_pressure}). Equating each radius to $r_{\rm acc}$ then gives:
\begin{equation}
\begin{aligned}
M_{\star,\rm PH}&=\frac{4\pi\alpha_{\rm B}}{3q}r_{\rm acc}^3
\left(\frac{n_0T_0}{T_{\rm ion}}\right)^2,\\
M_{\star,\rm DP}&=\frac{4\pi c n_0k_{\rm B}T_0}{f_{\rm abs}\ell}r_{\rm acc}^2.
\end{aligned}
\label{eq:radius_mass_thresholds}
\end{equation}
Here $q$ and $\ell$ are the source output per unit mass of stars formed and $f_{\rm abs}$ is the absorbed fraction. The zero-age \textsc{bpass} stellar populations for a median stellar metallicity of $Z=0.0018$ give $q\simeq7.8\times10^{46}\,{\rm s^{-1}\,M_\odot^{-1}}$ and $\ell\simeq4.9\times10^{36}\,{\rm erg\,s^{-1}\,M_\odot^{-1}}$. The threshold masses scale as $r_{\rm acc}^3n_0^2$ and $r_{\rm acc}^2n_0/f_{\rm abs}$ at fixed ambient temperatures. These cold-medium benchmark reference estimates omit gravity and inflow ram pressure as a source of confinement for the gas. A full absorption assumption ($f_{\rm abs} = 1$) yields the lowest direct-pressure mass threshold.

% \section{Some extra material}

% If you want to present additional material which would interrupt the flow of the main paper,
% it can be placed in an Appendix which appears after the list of references.

%%%%%%%%%%%%%%%%%%%%%%%%%%%%%%%%%%%%%%%%%%%%%%%%%%

% Don't change these lines
\bsp	% typesetting comment
\label{lastpage}
\end{document}